\documentclass{aastex701}

\usepackage{amsmath}
\usepackage[version=4]{mhchem}

\begin{document}

\title{Organosulfur Chemistry on sub-Neptunes:\\Implications for hazes and biosignatures}

\author[0000-0002-2828-0396]{Sean Jordan}
\affiliation{ETH Zurich, Institute for Particle Physics \& Astrophysics, Wolfgang-Pauli-Str. 27, 8093 Zurich, Switzerland}
\email[show]{jordans@ethz.ch}

\author[0000-0000-0000-0000]{Shang-Min Tsai}
\affiliation{Institute of Astronomy and Astrophysics (ASIAA), Taipei, Taiwan}
\email{smtsai@asiaa.sinica.edu.tw}

\author[0000-0002-7180-081X]{Paul B. Rimmer}
\affiliation{Cavendish Laboratory, University of Cambridge, JJ Thomson Ave, Cambridge CH3 0HE, UK}
\email{pbr27@cam.ac.uk}

\author[0000-0002-8713-1446]{Oliver Shorttle}
\affiliation{Institute of Astronomy, University of Cambridge, Cambridge CB3 0HA, UK}
\affiliation{Department of Earth Sciences, University of Cambridge, Cambridge CB2 3EQ, UK}
\email{os258@cam.ac.uk}

\begin{abstract}

The organosulfur biosignature gases dimethylsulfide (DMS) and dimethlydisulfide (DMDS) have recently been claimed to be present in the atmosphere of sub-Neptune exoplanet K2-18b, leading to the suggestion of possible extraterrestrial life.
Abiotic formation pathways for DMS and DMDS in reducing atmospheres have also been proposed, raising concern over the use of DMS and DMDS as biosignature gases more generally.
In this paper we independently test and contrast the proposed abiotic formation pathways for DMS and DMDS using K2-18b as a case study, and explore the wider implications for the atmospheric carbon and sulfur chemistry of hydrogen-rich sub-Neptunes.
We demonstrate that one proposed formation pathway is capable of producing observable abundances of abiotic DMS and DMDS, however it depends sensitively on the energy barrier of the limiting step, which remains unmeasured experimentally.
\textcolor{black}{The formation of} hydrocarbons including \ce{C2H6}, \textcolor{black}{however, occurs abundantly} and offers a plausible alternative explanation to the reported suggestions of organosulfur compounds on K2-18b, having previously been shown to \textcolor{black}{share similar spectral features with DMS and DMDS at near-IR wavelengths}.
Finally, we demonstrate that sulfur hazes form via the photochemistry of \ce{H2S} and condense in the atmosphere of K2-18b even at trace abundances.
We propose that variation in atmospheric sulfur abundance can explain the diversity of haziness observed across the sub-Neptune population so far with JWST.

\end{abstract}



\section{Introduction} 
\label{sec:Intro}

The James Webb Space Telescope (JWST) has revealed a variety of chemical features and haze layers across the sub-Neptune population. \textcolor{black}{In particular, methane (\ce{CH4}) and/or water (\ce{H2O}) have been robustly detected on a variety of the haze-free sub-Neptunes characterised by the James Webb Space Telescope (JWST) \citep[e.g.,][]{Davenport2025,Piaulet2024}. Additionally, \ce{CO2} has been detected on GJ3470b, TOI-270d, and K2-18b, \ce{CS2} on TOI-270d, \ce{SO2} on GJ3470b, and haze scattering has been detected on LP791-18c \citep{Beatty2024,Benneke2024,Hu2025,Roy2025}.} The origin of the detected molecules, and the underlying parameters controlling the variability of sub-Neptune haziness, are each currently under debate. Most notorious of these debates is the claim of the organosulfur compounds dimethylsulfide (\textcolor{black}{\ce{CH3SCH3};} DMS) and dimethyldisulfide (\textcolor{black}{\ce{CH3S2CH3};} DMDS) on the sub-Neptune exoplanet K2-18b \citep{Madhu2025}.

The organosulfur compounds DMS and DMDS are considered to be biosignature gases on exoplanets \citep[e.g.,][]{Pilcher2003,DomagalGoldman2011}. DMS and DMDS are produced primarily by marine biota on Earth \citep[e.g.,][]{Cooper1987, AnejaCooper1989} and were first suggested as exoplanet biosignature gases in recognition of the fact that the \ce{O2}-\ce{CH4} pair \citep{Lovelock1965} would not be relevant for tracing the microbial ecosystem of the early Earth \citep{Pilcher2003} prior to oxygenation of the atmosphere at 2.4\,Ga \citep{Holland2002}. Organosulfur compounds, in contrast, would have been produced biotically during the Archean by primitive terrestrial life during the degradation of one of life's universal amino acids---methionine \citep{KieneVisscher1987}---which continues to occur today as the modern Earth's primary organosulfur flux to the atmosphere \citep{Pilcher2003}.
Higher up in the Earth's atmosphere organosulfur molecules are susceptible to photochemical destruction by Sunlight and destruction by hydroxyl (\ce{OH}) radicals \citep{DomagalGoldman2011}, however within a more reducing atmosphere, and under the low-UV light of a quiescent M-dwarf star, organosulfur molecules could build up to potentially detectable abundances in an exoplanet's atmosphere \citep{DomagalGoldman2011, Seager2013, Tsai2024}. 

The first suggestion of an organosulfur biosignature gas being detected on an exoplanet has now been reported by \citet{Madhu2025}, not on an Earth-like exoplanet with an Archean-like atmosphere, but on the sub-Neptune exoplanet K2-18b \citep{Montet2015}.
K2-18b orbits an M-type host star and has a mass of 8.63$\pm1.35\,\mathrm{M}_{\oplus}$ and radius of 2.61$\pm0.09\,R_{\oplus}$ \citep{Cloutier2017,Benneke2017}. The observed mass and radius of K2-18b result in a mean density that is degenerate between possessing a magma ocean surface under a thick \ce{H2} atmosphere \citep{Shorttle2024, Rigby2024}, a water ocean surface under a thin \ce{H2} atmosphere \citep{Madhu2023}, or no well-defined surface at all in a deep supercritical envelope \citep{Wogan2024, Hu2025}. 
Efforts to distinguish between these hypothesised surface conditions and interior structures have resulted in six transits of K2-18b being captured by JWST so far \citep{Madhu2023, Madhu2025, Hu2025}. 
The transmission spectra provide evidence that K2-18b's upper atmosphere is dominated by \ce{H2}, with a chemical inventory of \ce{CH4} at a mixing ratio between $\sim$\,1\,\% to $\sim$\,10\,\% \citep{Madhu2023,Schmidt2025,Luque2025, Hu2025}, \ce{CO2} at a mixing ratio between $\sim$\,0.1\,\% to $\sim$\,1\,\% \citep{Hu2025,Luque2025}, and non-detections of \ce{NH3}, \ce{H2O} and \ce{H2S} \citep{Madhu2023,Schmidt2025,Luque2025,Lorenzo2025,Hu2025}, although \ce{H2O} has been inferred to be present in the hotter deep atmosphere from the observed \ce{CO2}:\ce{CH4} ratio \citep{Hu2025}.

The low significance detections ($<$\,3$\sigma$) of DMS and DMDS \citep{Madhu2023,Madhu2025} add to the observed and inferred chemical inventory of K2-18b. The initial announcement garnered extensive media attention due to the \textcolor{black}{possibility that these molecules may have had a biological origin} \citep{Madhu2025}. It has been highly disputed whether there is statistical evidence for DMS or DMDS in the transmission spectra of K2-18b \citep{Luque2025,MattLuis2025,Stevenson2025}, or whether it is physically plausible that K2-18b could possess a water ocean due to the challenge of forming with the large required water mass fraction \citep{Aaron2025}, and the difficulty of maintaining habitable conditions at the hypothetical ocean-atmosphere interface \citep{Innes2023, Leconte2024, Jordan2025b}. Nonetheless, following the tentative detections of DMS and DMDS, there is now a renewed interest in the use of organosulfur compounds as biosignature gases in non-Earth-like environments generally, particularly under the reducing conditions that prevail in \ce{H2}-dominated atmospheres where abiotic sources of organosulfur molecules may be more plausible.

Abiotic sources of DMS and DMDS have been discussed since at least as early as \citet{HeinenLauwers1996}, who found the formation of DMDS and various thiols, primarily methanethiol (\ce{CH3SH}), via aqueous reaction between \ce{FeS} and \ce{H2S} in the presence of \ce{CO2}. Since then, organosulfur compounds including DMS have been observed in both cometary matter \citep{Hanni2024} and the interstellar medium (ISM) \citep{Sans-Novo2025} suggesting that exogeneous delivery could supply organosulfur molecules to exoplanet atmospheres. Whether the exogenous delivery of organosulfur molecules poses a false positive biosignature in an exoplanet atmosphere would depend in detail on the astrophysical environment and efficiency of the delivery mechanism. 

Alternatively, organosulfur molecules could be produced abiotically \textit{in situ} in an exoplanet atmosphere via atmospheric chemistry. Previous experimental studies have demonstrated that organosulfur molecules and hydrocarbons can form naturally in gas mixtures containing \ce{CH4} and \ce{H2S} when subjected to elecric glow discharge or ultraviolet irradiation \citep[e.g.,][]{RaulinToupance1975,Reed2024}. In particular, laboratory photochemical experiments from \citet{Reed2024}, in which precursor species \ce{CH4} and \ce{H2S} were irradiated with ultraviolet light, produced significant abundances of DMS. \citet{Reed2024} propose a photochemically catalysed chemical pathway that may be occurring in the gas mixture:
\begin{align}
    \ce{H2S + h\nu} &\longrightarrow \ce{H + HS},\label{eq1}\\
    \ce{CH4 + h\nu} &\longrightarrow \ce{CH3 + H},\label{eq2}\\
    \ce{CH3 + HS + M} &\longrightarrow \ce{CH3SH + M},\label{eq3}\\
    \ce{CH3SH + h\nu} &\longrightarrow \ce{CH3S + H},\label{eq4}\\
    \ce{CH3S + CH3 + M} &\longrightarrow \ce{CH3SCH3 + M},\label{eq5}
\end{align}
where the \ce{CH3S} radicals, required for final DMS or DMDS formation, are produced photochemically. \textcolor{black}{\footnote{Note that \ce{HS} is equivalent to \ce{SH} and is the naming convention adopted here.}}

An alternative pathway for photochemically-catalysed DMS formation in the atmosphere of K2-18b was proposed by \citet{Hu2025} based on a reaction mechanism predicted by the Reaction Mechanism Generator \citep{Gao2016}. The pathway proceeds in a similar fashion to that suggested by \citet{Reed2024}, 
however the production of \ce{CH3S} radicals differs:

\begin{align}
    \ce{H2S + h\nu} &\longrightarrow \ce{H2 + S},\label{eq6}\\
    \ce{S + CH4} &\longrightarrow \ce{CH3 + HS},\label{eq7}\\
    \ce{S + CO + M} &\longrightarrow \ce{OCS + M},\label{eq8}\\
    \ce{CH3 + OCS} &\longrightarrow \ce{CH3S + CO},\label{eq9}\\
    \ce{CH3 + S + M} &\longrightarrow \ce{CH2SH + M},\label{eq10}\\
    \ce{CH3S + H + M} &\longrightarrow \ce{CH3SH + M},\label{eq11}\\
    \ce{CH2SH + CH3SH} &\longrightarrow \ce{CH3SH + CH3S},\label{eq12}\\
    \ce{CH3S + CH3 + M} &\longrightarrow \ce{CH3SCH3 + M}.\label{eq13}
\end{align}

Under this pathway, \ce{CH3S} radicals are not produced directly from photodissociation of \ce{CH3SH} but instead are formed through a catalytic hydrogen-abstraction reaction (reaction \ref{eq12}). Mechanistically, reaction \ref{eq12} proceeds with \ce{CH2SH} taking a hydrogen atom from a \ce{CH3SH} molecule, which reforms a new \ce{CH3SH} molecule and converts the original \ce{CH3SH} into a \ce{CH3S} radical \citep{Hu2025}. Stoichiometrically, \ce{CH3SH} thus behaves as a catalyst for the unimolecular conversion of \ce{CH2SH} to \ce{CH3S}, where the radical electron is on the carbon atom of \ce{CH2SH}, but on the sulfur atom of the \ce{CH3S}\,---\,the \ce{CH3S} radical must otherwise be produced directly via photochemistry under the proposed \citet{Reed2024} pathway. This catalytic H-abstraction reaction has no measured rate constant, with rate data estimated from the Reaction Mechanism Generator algorithm \citep{Hu2025, Gao2016}.

In this paper, we independently test the two proposed pathways in an extensive chemical network, and contrast their DMS and DMDS production efficiencies. We test whether the pathways can produce DMS or DMDS abiotically in the atmosphere of K2-18b as a function of the deep \ce{H2S} abundance and energy barrier to the hydrogen abstraction reaction suggested by \citet{Hu2025}. We investigate the wider carbon and sulfur chemistry on K2-18b and demonstrate that hydrocarbons \textcolor{black}{are always produced abundantly,} and sulfur hazes are produced \textcolor{black}{in correlation with the deep \ce{H2S} abundance}, pointing to the potential for organosulfur haze chemistry on sub-Neptunes generally. We end by discussing the implications for organosulfur biosignatures in reducing atmospheres and the potential prevalence of sulfur or organosulfur haze layers across the presently characteriseable population of sub-Neptunes accessible with JWST.

\section{The Chemical Network}

To independently test the abiotic formation pathways of DMS and DMDS in the atmosphere of K2-18b, we incorporate the reactions suggested by \citet{Reed2024} and \citet{Hu2025} into the chemical network \textcolor{black}{\textsc{Stand2024}} \citep[see][]{Rimmer2021,Jordan2025a,Tsai2023}. \textcolor{black}{The updated \textsc{Stand} network is then used to simulate the atmospheric chemistry with a coupled photochemical-kinetics model (see Appendix)}. \textsc{Stand} is a network of reactions assembled for photochemical-kinetics models of planetary and exoplanetary atmospheres \citep{Rimmer2016}. The chemical network is constructed to be holistic and applicable generally to the full compositional range of planet atmospheres, ranging from hot Jupiters \citep[e.g.,][]{Tsai2023} to sub-Neptunes \citep[e.g.,][]{Shorttle2024}, to terrestrial planets like the Earth, Venus, and their exoplanetary analogues \citep[e.g.,][]{Rimmer2019,Rimmer2021,Jordan2025a}. With the updates of Table \ref{table}, the reaction network \textsc{Stand} consists of 3588 chemical reactions: 178 unimolecular (photochemical reactions); 3045 bimolecular (two-body reactions); 365 termolecular (three-body reactions, with high and low pressure limits). With these reactions, \textsc{Stand} models 359 neutral and ionised species (196 neutral/163 ions) consisting of elements C/H/N/O/S up to chemical formulae of \ce{C5H8}/\ce{C10H18O2}/\ce{C2H6S2} with further elements and species also included which are not relevant to this study.

\begin{table}
\begin{center}
\caption{Bimolecular rate constants are calculated as: $k = \alpha ( \frac{ T}{ 300 K })^{\beta} e^{-\frac{\gamma}{T}}$. Termolecular rate constants are calculated for both the low pressure limit (first row data) and the high pressure limit (second row data) and then combined together as described in \citet{Rimmer2016}. Reactions with double arrows are reversed in \textsc{Stand}. Reactions with only a single arrow are not reversed because of the lack of the thermochemical NASA7 polynomial data for \ce{CH3SCH2} and \ce{CH2SH}. The termolecular reactions with the reference `Est.' do not have measured reaction rate data and are estimated here as rough upper limits close to the theoretical collisional limit so that the intermediate limiting steps to DMS and DMDS formation can be tested. \label{table}}
\begin{tabular}{l c c c r}
\\\hline\hline 
Reaction &  &  &  & Reference \\\hline\hline
\emph{Unimolecular} & \multicolumn{3}{c}{} & \\\hline
(R1) \;\ce{CH3SH + h\nu \longrightarrow CH3S + H} & \multicolumn{4}{r}{MPI-Mainz \citep{MPI-Mainz2013}} \\
(R2) \;\ce{CH3SCH3 + h\nu \longrightarrow CH3S + CH3} & \multicolumn{4}{r}{MPI-Mainz \citep{MPI-Mainz2013}}  \\
(R3) \;\ce{CH3S2CH3 + h\nu \longrightarrow CH3S + CH3S} & \multicolumn{4}{r}{MPI-Mainz \citep{MPI-Mainz2013}}  \\
(R4) \;\ce{CH3S2CH2 + h\nu \longrightarrow CH3S + CH2} & \multicolumn{4}{r}{MPI-Mainz \citep{MPI-Mainz2013}}  \\
\hline
\emph{Bimolecular} & $\alpha$ & $\beta$ & $\gamma$ &  \\\hline
(R5) \;\ce{CH3SCH3 + H <=> CH3 + CH3SH} & 2.84$\times10^{-11}$& 0.0 & 1.32$\times10^3$& \citet{Yokota1979} \\
(R6) \;\ce{CH3S2CH3 + H <=> CH3S + CH3SH} & 9.47$\times10^{-12}$ & 0.0 & 5.00$\times10^{1}$ & \citet{Ekwenchi1980} \\ 
(R7) \;\ce{CH3S + CO <=> CH3 + OCS} & 8.52$\times10^{-13}$ & 1.57 & 3.36$\times10^{3}$ & \citet{Tang2008} \\ 
(R8) \;\ce{CH3S + C <=> CH3 + CS} & 3.00$\times10^{-10}$ & 0.0 & 0.0 & \citet{Vidal2017} \\ 
(R9) \;\ce{CH3S + O2 <=> CH3 + SO2} & 6.07$\times10^{-8}$ & --3.80 & 6.19$\times10^{3}$ & \citet{Zhu2006} \\ 
(R10) \;\ce{CH3SH + H <=> CH3S + H2} & 4.42$\times10^{-12}$ & 1.73 & 4.96$\times10^{2}$ & \citet{Kerr2015} \\ 
(R11) \;\ce{CH3SH + H <=> CH4 + HS} & 9.48$\times10^{-13}$ & 1.98 & 8.32$\times10^{3}$ & \citet{Kerr2015} \\ 
(R12) \;\ce{CH3SH + O <=> CH3S + HO} & 2.22$\times10^{-12}$ & 1.82 & 4.03$\times10^{1}$ & \citet{Gersen2017} \\ 
(R13) \;\ce{CH3SH + HO <=> CH3S + H2O} & 5.16$\times10^{-13}$ & 1.77 & --8.50$\times10^{2}$ & \citet{Tsai2024} \\ 
(R14) \;\ce{CH3SH + HS <=> CH3S + H2S} & 1.99$\times10^{-10}$ & 0.0 & 2.98$\times10^{3}$ & \citet{Vijver2015} \\ 
(R15) \;\ce{CH3S + HS <=> CH3SH + S} & 6.06$\times10^{-15}$ & 4.45 & 3.48$\times10^{3}$ & Est. by \citet{Hu2025} \\  
(R16) \;\ce{CH3SCH3 + HO \longrightarrow CH3SCH2 + H2O} & 1.13$\times10^{-11}$ & 0.0 & 2.53$\times10^{2}$ & \citet{Atkinson1997} \\ 
(R17) \;\ce{CH3SCH3 + NO3 \longrightarrow CH3SCH2 + HNO3} & 1.90$\times10^{-13}$ & 0.0 & --5.20$\times10^{2}$ & \citet{Atkinson2004} \\ 
(R18) \;\ce{CH3SCH3 + H \longrightarrow CH3SCH2 + H2} & 8.44$\times10^{-12}$ & 1.72 & 2.21$\times10^{3}$ & \citet{Zhang2005} \\ 
(R19) \;\ce{CH3SCH3 + CH3 \longrightarrow CH3SCH2 + CH4} & 6.92$\times10^{-13}$ & 0.0 & 4.61$\times10^{3}$ & \citet{Arthur1976} \\ 
(R20) \;\ce{CH2SH + CH3SH \longrightarrow CH3SH + CH3S} & 2.71$\times10^{-15}$ & 3.06 & 3.02$\times10^{2}$ & Est. by \citet{Hu2025}\footnote{\textcolor{black}{The energy barrier to this reaction is varied as a sensitivity test in Section \ref{sec:K2-18b}.}} \\
\hline
\emph{Termolecular} & \multicolumn{3}{c}{} & \\\hline
(R21) \;\ce{CH3S + CH3 + M <=> CH3SCH3 + M} & 3.05$\times10^{-26}$ & 0.0  & 0.0 & Est. here \\
 & 1.00$\times10^{-10}$ & 0.0 & 0.0 & Est. here \\
(R22) \;\ce{CH3S + CH3S + M <=> CH3S2CH3 + M} & 3.05$\times10^{-26}$ & 0.0  & 0.0 & Est. here \\
 & 1.00$\times10^{-10}$ & 0.0 & 0.0 & Est. here \\
(R23) \;\ce{CH3S + H + M <=> CH3SH + M} & 1.36$\times10^{-10}$ & --3.79 & 3.86$\times10^{2}$ & Est. by \citet{Hu2025}\footnote{\textcolor{black}{These reaction rates were estimated within the pressure range 0.9\,mbar\,--\,98.3\,bar \citep{Hu2025} and may become unsuitable outside of that range.}} \\ 
 & 1.93$\times10^{-8}$ & --0.31 & 4.71$\times10^{2}$ & Est. by \citet{Hu2025}$^{\rm b}$ \\
(R24) \;\ce{CH3 + S + M \longrightarrow CH2SH + M} & 2.98$\times10^{-9}$ & --4.62 & 9.15$\times10^{2}$ & Est. by \citet{Hu2025}$^{\rm b}$ \\ 
 & 1.02$\times10^{-9}$ & 7.10$\times10^{-2}$ & 6.49$\times10^{1}$ & Est. by \citet{Hu2025}$^{\rm b}$ \\
\hline
\hline
\end{tabular}
\end{center}
\end{table}

We include the reactions from Table \ref{table} into \textsc{Stand2024} each in turn, in order to separately test the proposed reaction pathways to DMS and DMDS formation \citep{Reed2024, Hu2025}. The termolecular reactions in the final step of DMS or DMDS formation (Table 1, R21 and R22) do not have measured or computed reaction rate data. We here prescribe rough upper limits on their reaction rates close to the theoretical collisional limit, in order to test the intermediate reaction pathways. \textcolor{black}{Decreasing the rates of reactions R21 and R22 also decreases the efficiency of DMS and DMDS formation, and future work should be carried out to estimate the reaction rates of these steps in the reaction network. For the purposes of this study, we keep these reaction rate constants fixed in order to explore the kinetics of the intermediate pathways that have estimated reaction rate data}, in particular the catalytic H-abstraction reaction proposed by \citet{Hu2025} (Table 1, R20). After incorporating and testing the networks, we find that the net reaction pathways leading to DMS and DMDS formation in the \textsc{Stand} network have minor differences to the proposed pathways described in Section \ref{sec:Intro}. We list below the two pathways, which we now refer to as pathway A (following \citet{Reed2024}) and pathway B (following \citet{Hu2025}):

\begin{align}
    &\textrm{[A]}\nonumber \\ 
    \ce{H2S + h\nu} &\longrightarrow \ce{HS + H},\label{eq1}\\
   2( \ce{CH4 + C$_x$H$_y$} &\longrightarrow \ce{CH3 + C_$x$H$_{y+1}$}),\label{eq2}\\
    \ce{CH3 + HS + M} &\longrightarrow \ce{CH3SH + M},\label{eq3}\\
    \ce{CH3SH + h\nu} &\longrightarrow \ce{CH3S + H},\label{eq4}\\
    \ce{CH3S + CH3 + M} &\longrightarrow \ce{CH3SCH3 + M}.\label{eqDMS}\\
    \textrm{or }
    \ce{CH3S + CH3S + M} &\longrightarrow \ce{CH3S2CH3 + M}.\label{eqDMDS}
\end{align}

\begin{align}
    &\textrm{[B]} \nonumber \\
    2(\ce{H2S + h\nu} &\longrightarrow \ce{HS + H}),\label{eq6}\\
    \ce{HS + HS} &\longrightarrow \ce{H2S + S},\label{eq7}\\
    2( \ce{CH4 + C$_x$H$_y$} &\longrightarrow \ce{CH3 + C_$x$H$_{y+1}$}),\label{eq2}\\
    \ce{CH3 + S + M} &\longrightarrow \ce{CH2SH + M},\label{eq8}\\
    \ce{CH2SH + CH3SH} &\longrightarrow \ce{CH3SH + CH3S},\label{eq9}\\
    \ce{CH3S + CH3 + M} &\longrightarrow \ce{CH3SCH3 + M}.\label{eq5}\\
    \textrm{or }
    \ce{CH3S + CH3S + M} &\longrightarrow \ce{CH3S2CH3 + M}.\label{eq5}
\end{align}

\begin{figure}[htb!]
    \centering
    \includegraphics[width=0.9\columnwidth]{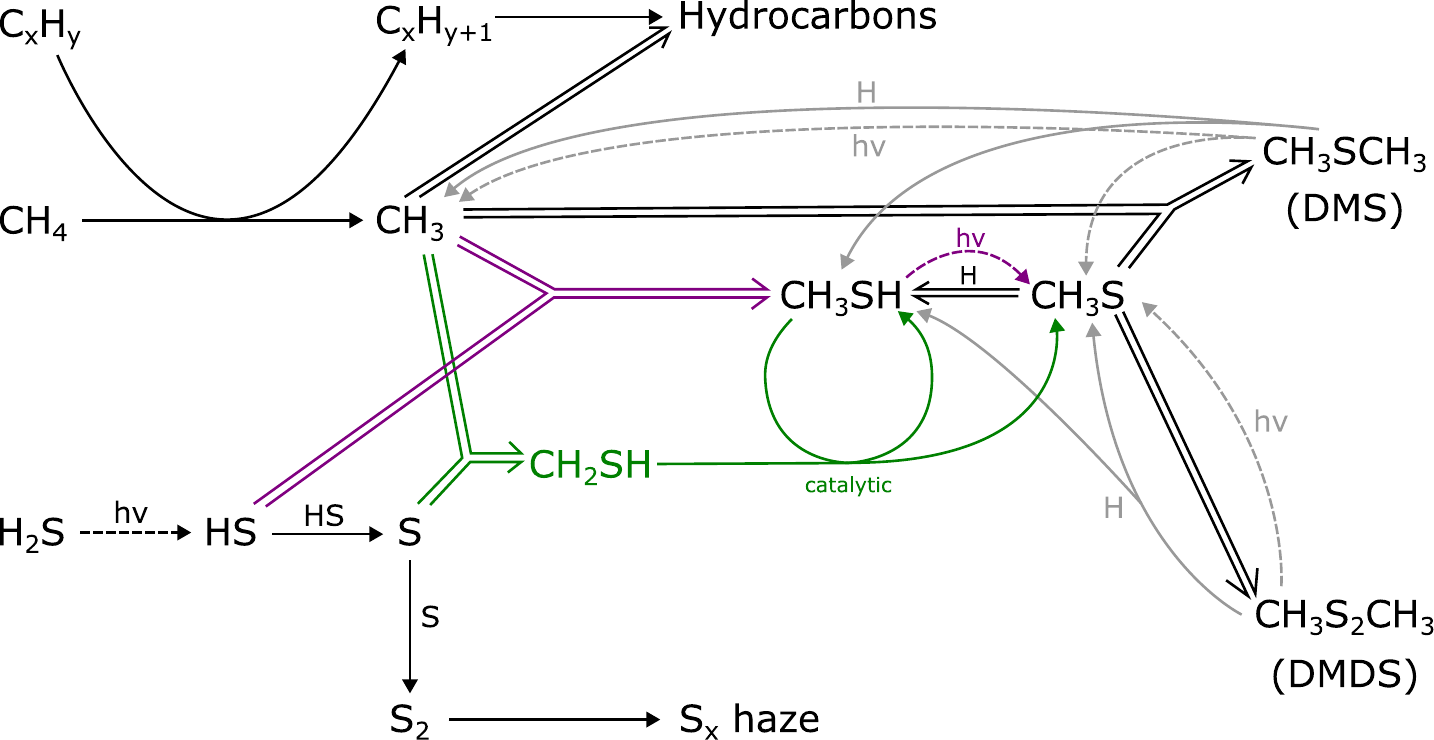}
    \caption{The network of important reactions for the carbon, sulfur, and coupled organosulfur chemistry in the atmosphere of K2-18b. Pathway B (following \citet{Hu2025}) dominates the DMS and DMDS production and the reactions unique to pathway B are highlighted in green. Reaction steps in pathway A (following \citet{Reed2024}) that are not important in pathway B are highlighted in purple. Destruction pathways of DMS and DMDS are highlighted in grey. Unimolecular reactions are denoted with dashed lines, bimolecular reactions with solid single lines, and termolecular reactions with solid double lines. \label{fig:network}}
\end{figure}

The final reaction network of important contributing reactions is shown in Figure \ref{fig:network}. 
\textcolor{black}{We now proceed by testing the efficiency of DMS and DMDS formation by pathway A compared to pathway B.} We test pathway A after including the relevant reactions for this pathway and before including the reactions for pathway B. We contrast this to results with the full network, including reactions for both pathways A and B, in which pathway B dominates.

\section{Abiotic DMS and DMDS production in Reducing Atmospheres}
\label{sec:compareDMS}

\subsection{\textcolor{black}{Comparing pathways A and B}}

DMS and DMDS can each be produced abiotically in a reduced atmosphere \textcolor{black}{by both pathways A and B}, only for relatively sluggish vertical mixing, becoming relatively inefficient for fast vertical mixing. In a 1\,bar \ce{H2}-atmosphere with a gas mixture of 10\%\,\ce{CH4} and varied \ce{H2S} mixing ratio, DMS and DMDS are produced photochemically in our chemical network for vertical mixing strengths below 10$^6$\,cm$^{2}$s$^{-1}$. For mixing faster than 10$^6\,$cm$^{2}$s$^{-1}$ the production of DMS and DMDS drops off rapidly, with the exception of DMDS production by pathway B under high \ce{H2S} conditions, which only drops off above 10$^7$\,cm$^{2}$s$^{-1}$. The drop in DMS and DMDS production occurs because the fast vertical mixing quenches \ce{H2S} to lower pressures which then self-shields its deeper photodissociation, allowing for DMS and DMDS production only in the uppermost atmosphere where DMS and DMDS are photochemically destroyed faster than they can be produced. The low values of vertical mixing required to sustain DMS and DMDS production are broadly consistent with the range of Earth-like or Venus-like vertical mixing parametrisations \citep{Rimmer2016,Rimmer2021}. Sluggish vertical mixing is also is an expected feature of atmospheres overlying magma-oceans due to convective-shutdown \citep{Harrison2025}.

\begin{figure}[htb!]
    \centering
    \includegraphics[width=0.7\columnwidth]{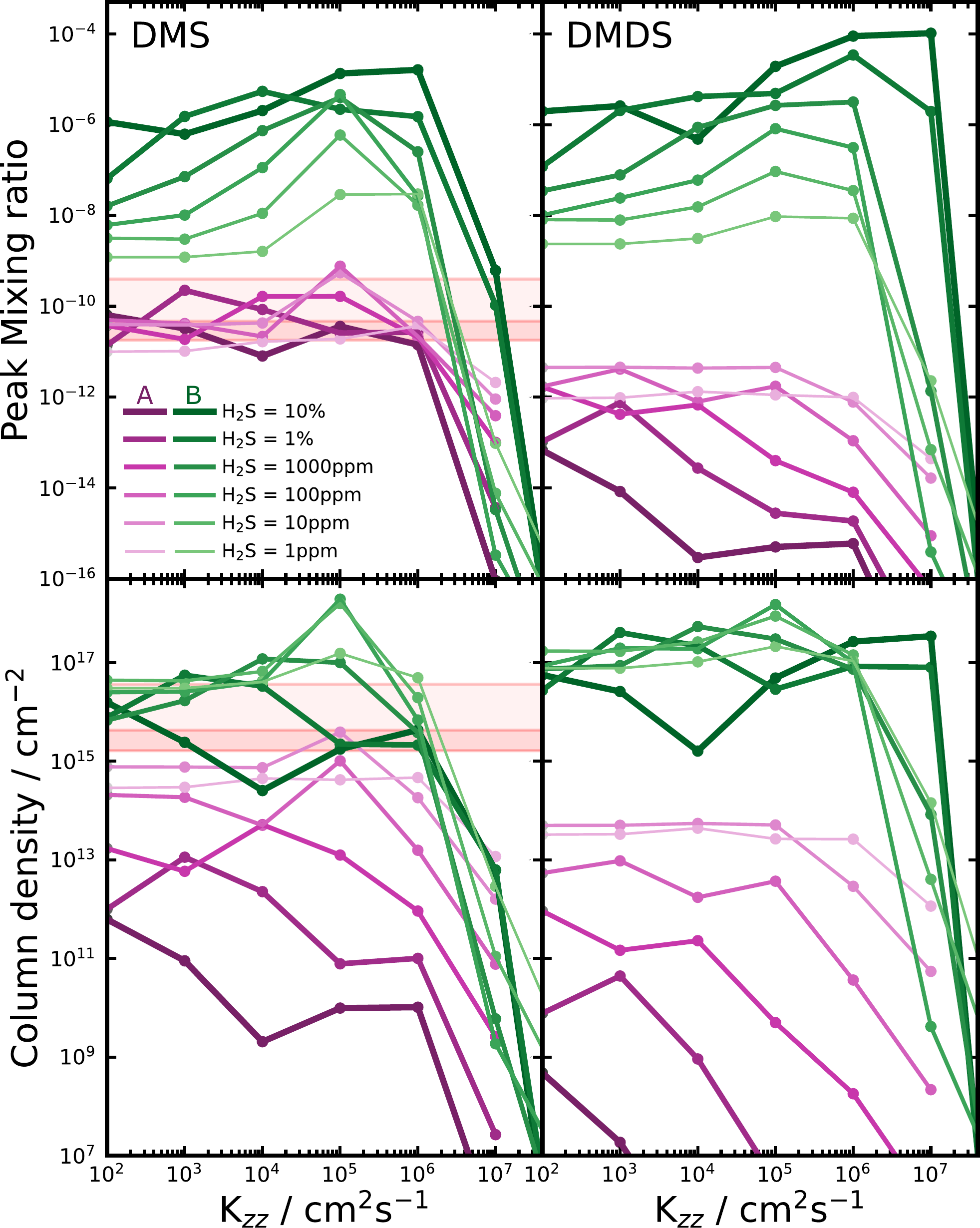}
    \caption{Maximum atmospheric \textcolor{black}{volume} mixing ratios (top) and integrated column number densities (bottom) of DMS (left) and DMDS (right) in a 1\,bar \ce{H2} atmosphere containing 10\% mole fraction of \ce{CH4} and varying \ce{H2S} abundances. The results of pathway B are highlighted in green and those of pathway A in purple. A range of reference DMS mixing ratios observed over \textcolor{black}{algae} blooms on Earth are highlighted in red \citep{Park2017} along with their equivalent integrated column density assuming uniform mixing ratio. \label{fig:DMS_DMDS}}
\end{figure}

For pathway A (Figure \ref{fig:DMS_DMDS}, \textcolor{black}{purple}) DMS is produced more abundantly than DMDS, and DMS is produced at comparable mixing ratios to those found on Earth over algal blooms (Figure \ref{fig:DMS_DMDS}, red shaded; \citet{Park2017}). The DMS mixing ratio is relatively insensitive to the input \ce{H2S} abundance: across a six order of magnitude difference in deep \ce{H2S} abundance, the DMS produced by pathway A, under sluggish vertical mixing, spans only 10$^{-11}$\,--\,10$^{-9}$ in mixing ratio. The peak of the photochemical production, however, occurs deeper in the atmosphere for lesser \ce{H2S} due to the self-shielding effect of photochemical reactions. The deeper \ce{H2S} photolysis produces \ce{HS} radicals that can react through pathway A to form \ce{CH3SH}, and ultimately DMS, faster under higher pressure due to the limiting termolecular step of getting from \ce{HS} to \ce{CH3SH} (reaction \ref{eq3}). Lower abundances of \ce{H2S} also results in less photochemical shielding of \ce{CH3SH}. These compounding effects lead to the unintuitive result that the equivalent column density of DMS produced by pathway A is anti-correlated with the \ce{H2S} abundance, down to \ce{H2S} $\sim$\,10\,ppm.

For pathway B (Figure \ref{fig:DMS_DMDS}, green) DMDS and DMS are both produced at abundances in excess of the Earth's biotic flux, and their mixing ratios positively correlate with the deep \ce{H2S} abundance. While the mixing ratios of DMS and DMDS correlate with the \ce{H2S} abundance, the column densities generally do not, particularly for \ce{H2S}\,$\lesssim$\,1\,\%. This reflects how pathway B, in contrast to pathway A, is not limited by a termolecular step (reaction \ref{eq3}) that proceeds significantly faster at higher pressure due the requirement of a third body in the reaction, but is instead limited by a bimolecular step (reaction \ref{eq9}), estimated by \citet{Hu2025} to be fast. This limiting step efficiently converts \ce{CH2SH}, formed from photochemical \ce{S} and thermochemical \ce{CH3} radicals, into \ce{CH3S}, which reacts with \ce{CH3} to form DMS or another \ce{CH3S} to form DMDS. The abiotic production of DMS and DMDS in reducing atmospheres is thus a possible abiotic false positive biosignature but depends crucially on the kinetics of the proposed pathway's limiting reaction. We now proceed by testing whether pathway B is capable of explaining the tentative detections of DMS or DMDS in the atmosphere of K2-18b, under the conditions from retrieval analyses, using self-consistent climate-chemistry modelling and varying the kinetics of the unmeasured limiting reaction (reaction \ref{eq9}). We additionally refer readers to the parallel study \citet{Tsai2026} where a general test of the abiotic production of DMS in the atmosphere of K2-18b has also been explored.

\subsection{DMS, DMDS, and hydrocarbon production on K2-18b}
\label{sec:K2-18b}

\begin{figure}[htb!]
    \centering
    \includegraphics[width=0.9\columnwidth]{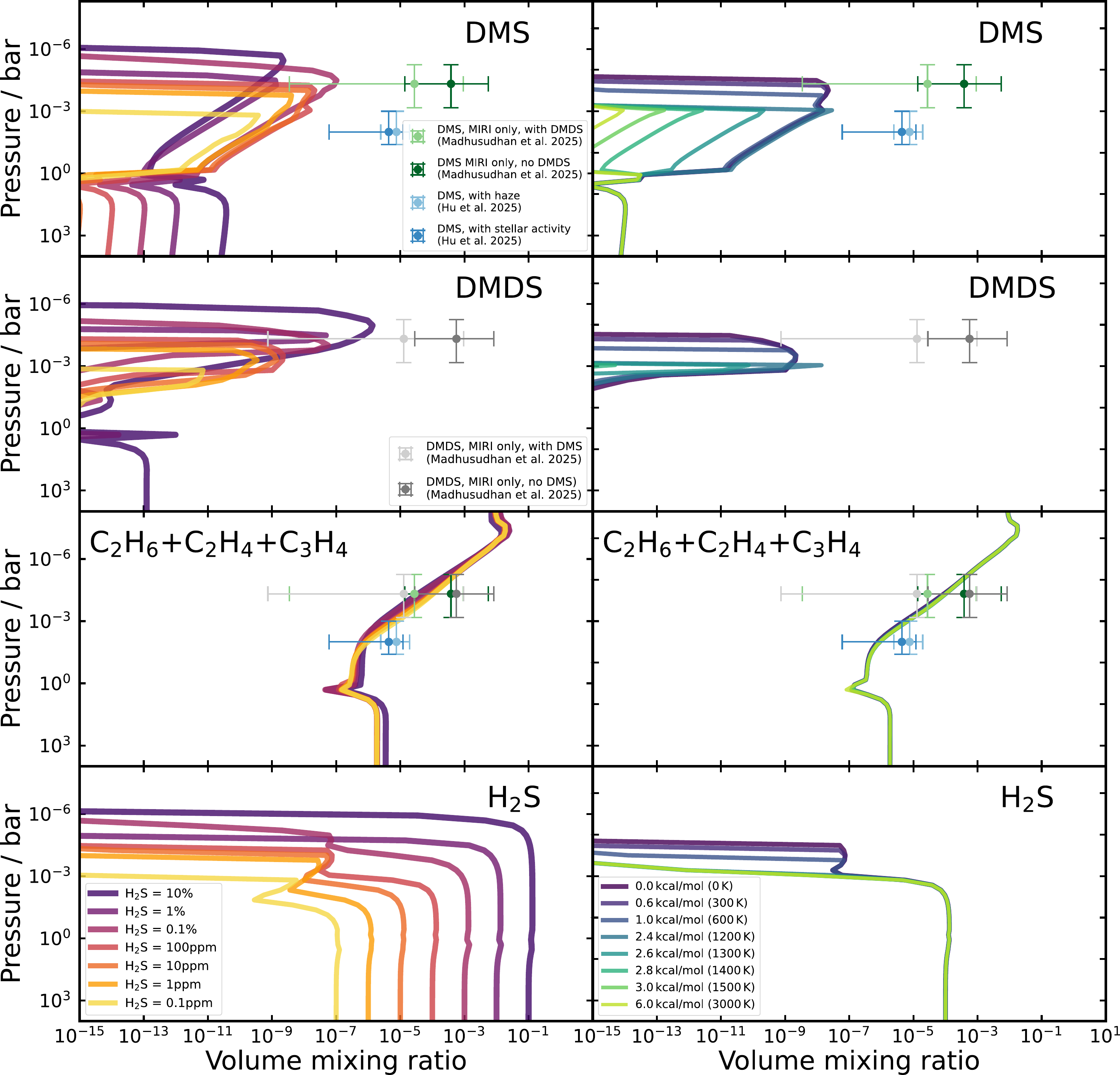}
    \caption{\textcolor{black}{Volume} mixing ratio profiles of DMS, DMDS, hydrocarbons \ce{C2H6}+\ce{C2H4}+\ce{C3H4}, and \ce{H2S}, in the atmosphere of K2-18b. Observational data points from \citet{Madhu2025} and \citet{Hu2025} are shown alongside, initially attributed to DMS and/or DMDS but later shown to be consistent with alternative hydrocarbon species \citep[e.g.,][]{Luque2025}. DMS and DMDS production is sensitive to the input \ce{H2S} abundance (left) and the energy barrier to reaction \ref{eq9} in pathway B (right), while the hydrocarbon production is not. \label{fig:K2-18b}}
\end{figure}

Under the retrieval constraints for the atmosphere of K2-18b \footnote{\textcolor{black}{$\sim$\,10\,\% \ce{CH4}, $\sim$\,0.1\,\%\,\ce{CO2}, and $\sim$\,25\,\%\,\ce{H2O} inferred in the deep atmosphere} \citep{Hu2025}}, abiotic production of DMS and DMDS \textcolor{black}{by pathway B} can satisfy the tentative reported detections from \citet{Madhu2025} within the lower bound of the 1\,$\sigma$ error threshold, if the deep \ce{H2S} abundance is $\gtrsim$\,10\,ppm (Figure \ref{fig:K2-18b}, left). The constraints are only satisfied when DMS and DMDS are considered in combination. In isolation, neither the DMS nor DMDS profiles can satisfy the detections from \citet{Madhu2025}, nor can the DMS profile satisfy the constraints reported from \citet{Hu2025}. 

The efficiency of abiotic production of DMS and DMDS is highly sensitive to the energy barrier of the catalytic H-abstraction reaction from \ce{CH3SH} to \ce{CH3S}. The activation energy barrier to this reaction was estimated in \citet{Hu2025} to be 0.6\,kcal\,mol$^{-1}$ \textcolor{black}{using an algorithm based on the kinetics of other H-abstraction reactions in the reaction mechanism generator package \citep{Gao2016}}. This low estimated energy barrier is comparable to the reaction proceeding with no energy barrier at all (Figure \ref{fig:K2-18b}, right), which is not necessarily unreasonable: the reaction is strongly exothermic with the H-S bond in methanethiol that is being broken requiring 368\,kJ\,mol$^{-1}$ while the H-C bond formed in mercaptan releasing 386\,kJ\,mol$^{-1}$ \citep{blanskyBONDS}. Exothermic reactions, nonetheless, are unlikely to proceed with no energy barrier and therefore we explore a range of increasing activation energy barriers until the abiotic production of DMS and DMDS becomes negligible. Between 2.4\,kcal\,mol$^{-1}$\,--\,3.0\,kcal\,mol$^{-1}$ the production of abiotic DMS or DMDS diminishes rapidly until $\gtrsim$\,3.0\,kcal\,mol$^{-1}$ they are diminished to non-detectable abundances. Whether or not the photochemically-catalysed abiotic formation of DMS and DMDS could pose a false positive bisignature in a reducing atmosphere like that of K2-18b, thus depends crucially on the reaction rate of the intermediate H-abstraction reaction from \citet{Hu2025}. Future work to calculate this reaction rate from quantum chemical calculations would be valuable.

In contrast, the hydrocarbons \textcolor{black}{\ce{C2H6}, \ce{C2H4}, and \ce{C3H4} would naturally satisfy all abundance constraints otherwise attributed to DMS and DMDS by both \citet{Madhu2025} and \citet{Hu2025}}. It has previously been demonstrated that hydrocarbons such as \ce{C2H6} \citep{Luque2025}, \textcolor{black}{\ce{C2H4}} \citep{Stevenson2025}, or \ce{C3H4} \citep{MattLuis2025} can fit the observed transmission spectra \textcolor{black}{of K2-18b equally well compared to} DMS or DMDS. Many hydrocarbon molecules in general are difficult to distinguish from DMS or DMDS in low resolution observations at JWST wavelengths \citep[see ][]{Niraula2025}. 
The \textcolor{black}{rate of hydrocarbon production is insensitive to the deep \ce{H2S} abundance or the energy barrier to the catalytic reaction step in pathway B (Figure \ref{fig:K2-18b})}. Our profile of hydrocarbons therefore provide a plausible and self-consistent alternative explanation to the reported detections of DMS or DMDS from both \citet{Madhu2023} and \citet{Hu2025}. We also refer readers to the parallel study \citet{Tsai2026} for further discussions on the hydrocarbon production on K2-18b, and spectral diagnostics of DMS and DMDS.

\citet{DomagalGoldman2011} suggested that higher-than-expected ethane production in an exoplanet atmosphere may be a proxy for biotic DMS, thanks to the enhanced supply of methyl radicals provided by DMS photolysis. Our results implicitly test this hypothesis and demonstrate that ethane overproduction by biotic DMS photolysis must reach mixing ratios significantly greater than 10$^{-5}$\,--\,$10^{-3}$ within observable pressure levels in order to overcome the abiotic baseline of ethane production. This constraint is demonstrated for 10\% \ce{CH4} in the atmosphere however, in an atmosphere with lesser \ce{CH4}, this ethane overproduction requirement may become less strict.

\section{A Sulfur haze deck on Sub-Neptunes}
\label{sec:haze}

\begin{figure}[htb!]
    \centering
    \includegraphics[width=0.9\columnwidth]{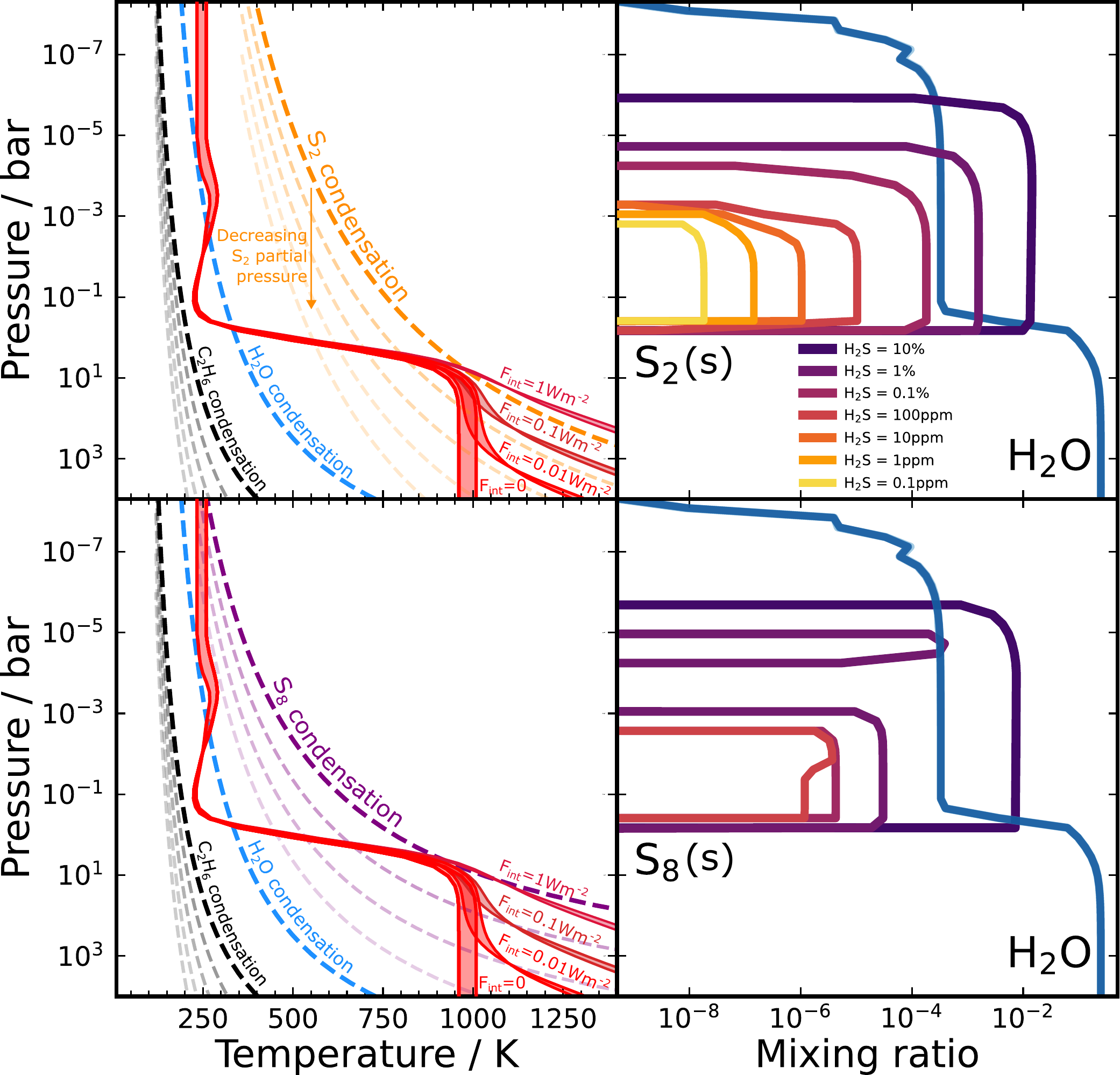}
    \caption{(Left) The temperature structure of K2-18b across the range of input \ce{H2S} abundances. The Antoine equations for the saturation vapour pressure of sulfur allotropes, \ce{H2O}, and \ce{C2H6} are shown in bold dashed lines, and in lighter coloured lines the vapour pressure curves are progressively scaled down from 100\% to 100\,ppm to illustrate how sulfur allotropes still condense out at trace mixing ratios. (Right) Profiles of gaseous \ce{H2O} and condensed \ce{S2} and \ce{S8} in the atmosphere of K2-18b. \textcolor{black}{For \ce{H2O} we show only the profile for the 100\,ppm \ce{H2S} model, and the differences to the \ce{H2O} condensation for the other models are insignificant.} \label{fig:haze}}
\end{figure}

The formation of organosulfur molecules from \ce{H2S} is only the secondary secondary consequence of \ce{H2S} photochemistry: the primary consequence of \ce{H2S} photochemistry in the atmosphere of K2-18b is the production of sulfur allotropes \citep{MajaPaul2025,Tsai2024}. \ce{H2S} is photodissociated to \ce{HS} radicals, and when two \ce{HS} radicals recombine to reform one \ce{H2S} molecule, they leave behind elemental \ce{S}. Sulfur atoms are thus produced in abundance around the pressure level of \ce{H2S} photolysis in the atmosphere, and react together to form sulfur allotropes (\ce{S_x} for $x$\,=\,2\,--\,8). Sulfur allotropes can condense at relatively high temperatures compared to water or hydrocarbons (Figure \ref{fig:haze}, left) and therefore we expect a sulfur allotrope haze to be formed on K2-18b if \ce{H2S} is present in the deep atmosphere, as has been previously predicted for giant exoplanet atmospheres \citep{Gao2017} and anoxic terrestrial atmospheres \citep{HuSeagerBains2013}. Since this sulfur haze forms where \ce{H2S} photodissociates, and immediately condenses as a photochemical haze, this can potentially block the deep \ce{H2S} from view observationally. This effect may be responsible for the non-detection of \ce{H2S} in transmission spectroscopy observations of the sub-Neptune atmospheres obtained so far with JWST, despite \ce{H2S} being the expected dominant carrier of deep atmospheric sulfur. 

Due to the high condensation temperatures of sulfur allotropes, sulfur hazes may be widespread in the sub-Neptune population and could be responsible for the hazy atmospheres that have been observed with JWST \citep[e.g.,][]{Roy2025}. Sulfur allotropes can react to grow larger in the condensed phase as well as in the gas phase, and the condensation temperatures of sulfur allotropes become lower as sulfur allotropes become longer. If a sulfur haze on K2-18b condenses as \ce{S2} then the formation of a haze deck is easily achieved across a wide variety of deep \ce{H2S} abundances. If however sulfur allotropes grow to \ce{S8}, the condensation depends more sensitively on the stratospheric temperature profile and is favoured by high \ce{H2S} abundances in order to produce enough \ce{S8} to be oversaturated. Figure \ref{fig:haze} demonstrates the two end-member scenarios in sulfur haze condensation: (Figure \ref{fig:haze}, top) all allotropes are allowed to condense, resulting in \ce{S2} immediately condensing out after photochemical formation; (Figure \ref{fig:haze}, bottom) condensation is turned off for all allotropes except \ce{S8}, however \ce{S8} must then be synthesised exclusively by gas phase chemistry in our model. Reactions between sulfur allotropes are generally very fast and haze condensation and evaporation also depends in detail on condensation nuclei and atmospheric dynamics. The true distribution of elemental sulfur in the atmosphere may therefore contain a mixture of shorter and longer allotropes across gaseous and condensed phases.

Due to the low condensation temperatures of hydrocarbons up to ethane, light hydrocarbons do not condense in the atmosphere of K2-18b.  Light hydrocarbons are thus also unlikely to be responsible for the haziness of  other characterised sub-Neptunes to date, where the equilibrium temperatures are greater and methane abundances lesser than in the case of K2-18b. Longer hydrocarbons will possibly be synthesised on sub-Neptunes and, due to their higher condensation temperatures, may be capable of forming soot-like hazes \citep{Kevin2009,Horst2018,Kawashima2018}. The synthesis of long chain hydrocarbons, however, is found to be inhibited in the presence of hydrogen from experimental haze studies: hydrogen addition will fill up the electron shell of carbon atoms of reactive methyl groups, truncating further reaction with other methyl groups in the hydrocarbon synthesis \citep{Dewitt2009}. In contrast, the presence of trace sulfur gases in haze experiments greatly enhances the haze production rate, and thus observed hazy sub-Neptunes possibly contain neither pure sulfur nor pure hydrocarbon hazes, but mixed-composition organo-sulfur hazes \citep{Louis2025}. The hydrocarbon chemistry and the sulfur chemistry may be tightly coupled on sub-Neptunes, even with only trace \ce{H2S} compared to \ce{CH4} abundances, and further observations with JWST may reveal whether sulfur or organosulfur hazes form generally across the sub-Neptune population.

\section{Discussion and conclusions}

The formation of DMS and DMDS abiotically on K2-18b and in reducing atmospheres generally depends on reactions without measured or computed reaction rate data. In particular, the activation energy for the catalytic H-abstraction reaction proposed in \citet{Hu2025} poses a very sensitive control on whether observable abundances of DMS can form in K2-18b's upper atmosphere. Quantum chemical calculations of this reaction will be crucial to ascertain if DMS is still to be considered a biosignature gas in sub-Neptune or Archean Earth-like exoplanet atmospheres \citep{Pilcher2003, DomagalGoldman2011, Seager2013, Eddie2024, Leung2025}.

\textcolor{black}{The short hydrocarbons \ce{C2H6}, \ce{C2H4}, and \ce{C3H4} all form in the atmosphere of K2-18b and can provide a self-consistent alternative explanation to the reported detections of DMS or DMDS from both \citet{Madhu2023} and \citet{Hu2025}. Previous studies have shown that these hydrocarbon gases are each capable of fitting the observed transmission spectrum of K2-18b equally well compared to DMS or DMDS \citep{Luque2025,Stevenson2025,MattLuis2025}. We demonstrate here that such hydrocarbon gases form abundantly within the atmospheric pressure levels probed by \citet{Madhu2025} and \citet{Hu2025} as a natural consequence of the high \ce{CH4} abundance (see Figure \ref{fig:K2-18b}). Our results may suggest that abiotic DMS and DMDS on sub-Neptunes in general will be masked by the similar absorption features of more abundant hydrocarbon species that form alongside DMS and DMDS in high-\ce{CH4} environments. The challenge of distinguishing myriad hydrocarbon species from DMS or DMDS \citep[see ][]{Niraula2025} in the wavelength range probed by JWST points to the requirement for more accurate cross sections of complex hydrocarbons and organosulfur species.}

We predict that elemental sulfur hazes are more easily formed than hydrocarbon hazes in K2-18b and sub-Neptunes generally, and are potentially responsible for the haze layers observed across the sub-Neptune population so far \citep{Davenport2025,Roy2025}. The propensity for hydrocarbon haze production to be shut down in \ce{H2}-atmospheres, but greatly enhanced by the presence of sulfur \citep[e.g.,][]{Tsai2024}, may suggest that sulfur and/or organosulfur hazes should be widespread in the sub-Neptune population. Experimental studies of organosulfur haze formation pathways will be crucial for the community to explore in the JWST era to interpret sub-Neptune transit spectra and future observations of early Earth analogues \citep{Louis2024}.

The sulfur abundance on temperate sub-Neptunes may be intrinsically difficult to constrain due to the photochemical product of \ce{H2S} being a condensible haze, however it may be possible in the context of photochemical modelling: at high abundances, \ce{H2S} can self-shield and may remain potentially observable above a cloud or haze deck; at lower abundances, the pressure level of the haze-top may implicitly trace the \ce{H2S} abundance because it is dependent on the effective \ce{H2S} UV-photosphere. The diversity in observed haze-top pressure levels constrained across K2-18b, TOI-270d, and LP591-18b, may indicate a diversity not only in the stellar-UV but also compositional diversity in the bulk sulfur abundance \citep{Roy2025}. 
While the observational opportunity afforded by the large scale heights of \ce{H2}-rich sub-Neptunes is an asset, at the population level, flat transmission spectra that are truncated by hazes will also play a pivotal role in unveiling the puzzle of haze formation and atmospheric composition across sub-Neptunes.

\section*{Acknowledgements}

\textcolor{black}{We thank our reviewer for improving the quality of this article}. S.J. acknowledges funding support from ETH Zurich and the NOMIS Foundation in the form of a research fellowship. The NOMIS Foundation ETH Fellowship Programme and respective research are made possible thanks to the support of the NOMIS Foundation. 
S-.M.T is supported by the National Science and Technology Council (grants 114-2112-M-001-065-MY3) and an Academia Sinica Career Development Award (AS-CDA-115-M03). O.S. acknowledges support from UKRI(STFC) grant UKRI1184. PBR received support from the Leverhulme Centre for Life in the Universe Joint Collaborations Research Project Grant G112026, Project KKZA/237.

\bibliography{refs.bib}{}

@ARTICLE{Montet2015,
       author = {{Montet}, Benjamin T. and {Morton}, Timothy D. and {Foreman-Mackey}, Daniel and {Johnson}, John Asher and {Hogg}, David W. and {Bowler}, Brendan P. and {Latham}, David W. and {Bieryla}, Allyson and {Mann}, Andrew W.},
        title = "{Stellar and Planetary Properties of K2 Campaign 1 Candidates and Validation of 17 Planets, Including a Planet Receiving Earth-like Insolation}",
      journal = {\apj},
         year = 2015,
        month = aug,
       volume = {809},
       number = {1},
          eid = {25},
        pages = {25},
          doi = {10.1088/0004-637X/809/1/25},
archivePrefix = {arXiv},
       eprint = {1503.07866},
 primaryClass = {astro-ph.EP},
       adsurl = {https://ui.adsabs.harvard.edu/abs/2015ApJ...809...25M}
}

@ARTICLE{Cloutier2017,
       author = {{Cloutier}, R. and {Astudillo-Defru}, N. and {Doyon}, R. and {Bonfils}, X. and {Almenara}, J.-M. and {Benneke}, B. and {Bouchy}, F. and {Delfosse}, X. and {Ehrenreich}, D. and {Forveille}, T. and {Lovis}, C. and {Mayor}, M. and {Menou}, K. and {Murgas}, F. and {Pepe}, F. and {Rowe}, J. and {Santos}, N.~C. and {Udry}, S. and {W{\"u}nsche}, A.},
        title = "{Characterization of the K2-18 multi-planetary system with HARPS. A habitable zone super-Earth and discovery of a second, warm super-Earth on a non-coplanar orbit}",
      journal = {\aap},
         year = 2017,
        month = dec,
       volume = {608},
          eid = {A35},
        pages = {A35},
          doi = {10.1051/0004-6361/201731558},
archivePrefix = {arXiv},
       eprint = {1707.04292},
 primaryClass = {astro-ph.EP},
       adsurl = {https://ui.adsabs.harvard.edu/abs/2017A&A...608A..35C}
}

@ARTICLE{Benneke2017,
       author = {{Benneke}, Bj{\"o}rn and {Werner}, Michael and {Petigura}, Erik and {Knutson}, Heather and {Dressing}, Courtney and {Crossfield}, Ian J.~M. and {Schlieder}, Joshua E. and {Livingston}, John and {Beichman}, Charles and {Christiansen}, Jessie and {Krick}, Jessica and {Gorjian}, Varoujan and {Howard}, Andrew W. and {Sinukoff}, Evan and {Ciardi}, David R. and {Akeson}, Rachel L.},
        title = "{Spitzer Observations Confirm and Rescue the Habitable-zone Super-Earth K2-18b for Future Characterization}",
      journal = {\apj},
         year = 2017,
        month = jan,
       volume = {834},
       number = {2},
          eid = {187},
        pages = {187},
          doi = {10.3847/1538-4357/834/2/187},
archivePrefix = {arXiv},
       eprint = {1610.07249},
 primaryClass = {astro-ph.EP},
       adsurl = {https://ui.adsabs.harvard.edu/abs/2017ApJ...834..187B}
}

@ARTICLE{Shorttle2024,
       author = {{Shorttle}, Oliver and {Jordan}, Sean and {Nicholls}, Harrison and {Lichtenberg}, Tim and {Bower}, Dan J.},
        title = "{Distinguishing Oceans of Water from Magma on Mini-Neptune K2-18b}",
      journal = {\apjl},
         year = 2024,
        month = feb,
       volume = {962},
       number = {1},
          eid = {L8},
        pages = {L8},
          doi = {10.3847/2041-8213/ad206e},
archivePrefix = {arXiv},
       eprint = {2401.05864},
 primaryClass = {astro-ph.EP},
       adsurl = {https://ui.adsabs.harvard.edu/abs/2024ApJ...962L...8S}
}

@ARTICLE{Madhu2023,
       author = {{Madhusudhan}, Nikku and {Sarkar}, Subhajit and {Constantinou}, Savvas and {Holmberg}, M{\r{a}}ns and {Piette}, Anjali A.~A. and {Moses}, Julianne I.},
        title = "{Carbon-bearing Molecules in a Possible Hycean Atmosphere}",
      journal = {\apjl},
         year = 2023,
        month = oct,
       volume = {956},
       number = {1},
          eid = {L13},
        pages = {L13},
          doi = {10.3847/2041-8213/acf577},
archivePrefix = {arXiv},
       eprint = {2309.05566},
 primaryClass = {astro-ph.EP},
       adsurl = {https://ui.adsabs.harvard.edu/abs/2023ApJ...956L..13M}
}

@ARTICLE{Wogan2024,
       author = {{Wogan}, Nicholas F. and {Batalha}, Natasha E. and {Zahnle}, Kevin J. and {Krissansen-Totton}, Joshua and {Tsai}, Shang-Min and {Hu}, Renyu},
        title = "{JWST Observations of K2-18b Can Be Explained by a Gas-rich Mini-Neptune with No Habitable Surface}",
      journal = {\apjl},
         year = 2024,
        month = mar,
       volume = {963},
       number = {1},
          eid = {L7},
        pages = {L7},
          doi = {10.3847/2041-8213/ad2616},
archivePrefix = {arXiv},
       eprint = {2401.11082},
 primaryClass = {astro-ph.EP},
       adsurl = {https://ui.adsabs.harvard.edu/abs/2024ApJ...963L...7W}
}

@ARTICLE{Hu2025,
       author = {{Hu}, Renyu and {Bello-Arufe}, Aaron and {Tokadjian}, Armen and {Yang}, Jeehyun and {Damiano}, Mario and {Roy}, Pierre-Alexis and {Coulombe}, Louis-Philippe and {Madhusudhan}, Nikku and {Constantinou}, Savvas and {Benneke}, Bj{\"o}rn},
        title = "{A water-rich interior in the temperate sub-Neptune K2-18 b revealed by JWST}",
      journal = {arXiv e-prints},
         year = 2025,
        month = jul,
          eid = {arXiv:2507.12622},
        pages = {arXiv:2507.12622},
          doi = {10.48550/arXiv.2507.12622},
archivePrefix = {arXiv},
       eprint = {2507.12622},
 primaryClass = {astro-ph.EP},
       adsurl = {https://ui.adsabs.harvard.edu/abs/2025arXiv250712622H}
}

@ARTICLE{Rigby2024,
       author = {{Rigby}, Frances E. and {Pica-Ciamarra}, Lorenzo and {Holmberg}, M{\r{a}}ns and {Madhusudhan}, Nikku and {Constantinou}, Savvas and {Schaefer}, Laura and {Deng}, Jie and {Lee}, Kanani K.~M. and {Moses}, Julianne I.},
        title = "{Toward a Self-consistent Evaluation of Gas Dwarf Scenarios for Temperate Sub-Neptunes}",
      journal = {\apj},
         year = 2024,
        month = nov,
       volume = {975},
       number = {1},
          eid = {101},
        pages = {101},
          doi = {10.3847/1538-4357/ad6c38},
archivePrefix = {arXiv},
       eprint = {2409.03683},
 primaryClass = {astro-ph.EP},
       adsurl = {https://ui.adsabs.harvard.edu/abs/2024ApJ...975..101R}
}

@ARTICLE{Madhu2025,
       author = {{Madhusudhan}, Nikku and {Constantinou}, Savvas and {Holmberg}, M{\r{a}}ns and {Sarkar}, Subhajit and {Piette}, Anjali A.~A. and {Moses}, Julianne I.},
        title = "{New Constraints on DMS and DMDS in the Atmosphere of K2-18 b from JWST MIRI}",
      journal = {\apjl},
         year = 2025,
        month = apr,
       volume = {983},
       number = {2},
          eid = {L40},
        pages = {L40},
          doi = {10.3847/2041-8213/adc1c8},
archivePrefix = {arXiv},
       eprint = {2504.12267},
 primaryClass = {astro-ph.EP},
       adsurl = {https://ui.adsabs.harvard.edu/abs/2025ApJ...983L..40M}
}

@ARTICLE{Luque2025,
       author = {{Luque}, R. and {Piaulet-Ghorayeb}, C. and {Radica}, M. and {Xue}, Q. and {Zhang}, M. and {Bean}, J.~L. and {Samra}, D. and {Steinrueck}, M.~E.},
        title = "{Insufficient evidence for DMS and DMDS in the atmosphere of K2-18 b: From a joint analysis of JWST NIRISS, NIRSpec, and MIRI observations}",
      journal = {\aap},
         year = 2025,
        month = aug,
       volume = {700},
          eid = {A284},
        pages = {A284},
          doi = {10.1051/0004-6361/202555580},
archivePrefix = {arXiv},
       eprint = {2505.13407},
 primaryClass = {astro-ph.EP},
       adsurl = {https://ui.adsabs.harvard.edu/abs/2025A&A...700A.284L}
}

@ARTICLE{MattLuis2025,
       author = {{Welbanks}, Luis and {Nixon}, Matthew C. and {McGill}, Peter and {Tilke}, Lana J. and {Wiser}, Lindsey S. and {Rotman}, Yoav and {Mukherjee}, Sagnick and {Feinstein}, Adina and {Line}, Michael R. and {Benneke}, Bj{\"o}rn and {Seager}, Sara and {Beatty}, Thomas G. and {Seligman}, Darryl Z. and {Parmentier}, Vivien and {Sing}, David},
        title = "{The Challenges Involved in the Detection of Gases in Exoplanet Atmospheres}",
      journal = {arXiv e-prints},
         year = 2025,
        month = apr,
          eid = {arXiv:2504.21788},
        pages = {arXiv:2504.21788},
          doi = {10.48550/arXiv.2504.21788},
archivePrefix = {arXiv},
       eprint = {2504.21788},
 primaryClass = {astro-ph.EP},
       adsurl = {https://ui.adsabs.harvard.edu/abs/2025arXiv250421788W}
}

@ARTICLE{Lorenzo2025,
       author = {{Pica-Ciamarra}, Lorenzo and {Madhusudhan}, Nikku and {Cooke}, Gregory J. and {Constantinou}, Savvas and {Binet}, Martin},
        title = "{A Systematic Search for Trace Molecules in Exoplanet K2-18 b}",
      journal = {arXiv e-prints},
         year = 2025,
        month = may,
          eid = {arXiv:2505.10539},
        pages = {arXiv:2505.10539},
          doi = {10.48550/arXiv.2505.10539},
archivePrefix = {arXiv},
       eprint = {2505.10539},
 primaryClass = {astro-ph.EP},
       adsurl = {https://ui.adsabs.harvard.edu/abs/2025arXiv250510539P}
}

@ARTICLE{Schmidt2025,
       author = {{Schmidt}, Stephen P. and {MacDonald}, Ryan J. and {Tsai}, Shang-Min and {Radica}, Michael and {Wang}, Le-Chris and {Ahrer}, Eva-Maria and {Bell}, Taylor J. and {Fisher}, Chloe and {Thorngren}, Daniel P. and {Wogan}, Nicholas and {May}, Erin M. and {Ferrari}, Piero and {Bennett}, Katherine A. and {Rustamkulov}, Zafar and {L{\'o}pez-Morales}, Mercedes and {Sing}, David K.},
        title = "{A Comprehensive Reanalysis of K2-18 b's JWST NIRISS+NIRSpec Transmission Spectrum}",
      journal = {\aj},
         year = 2025,
        month = dec,
       volume = {170},
       number = {6},
          eid = {298},
        pages = {298},
          doi = {10.3847/1538-3881/ae019a},
archivePrefix = {arXiv},
       eprint = {2501.18477},
 primaryClass = {astro-ph.EP},
       adsurl = {https://ui.adsabs.harvard.edu/abs/2025AJ....170..298S}
}

@ARTICLE{MajaPaul2025,
       author = {{Radecka}, Maja W. and {Rimmer}, Paul B.},
        title = "{Nitrogen chemistry of hycean worlds on the example of K2-18b}",
      journal = {\mnras},
         year = 2025,
        month = oct,
       volume = {543},
       number = {1},
        pages = {789-812},
          doi = {10.1093/mnras/staf1476},
archivePrefix = {arXiv},
       eprint = {2509.03455},
 primaryClass = {astro-ph.EP},
       adsurl = {https://ui.adsabs.harvard.edu/abs/2025MNRAS.543..789R}
}

@ARTICLE{Stevenson2025,
       author = {{Stevenson}, Kevin B. and {Lustig-Yaeger}, Jacob and {May}, E.~M. and {Kopparapu}, Ravi K. and {Fauchez}, Thomas J. and {Haqq-Misra}, Jacob and {Limbach}, Mary Anne and {Schwieterman}, Edward W. and {Sotzen}, Kristin S. and {Tsai}, Shang-Min},
        title = "{K2-18b Does Not Meet the Standards of Evidence for Life}",
      journal = {\aj},
         year = 2025,
        month = nov,
       volume = {170},
       number = {5},
          eid = {257},
        pages = {257},
          doi = {10.3847/1538-3881/ae0338},
archivePrefix = {arXiv},
       eprint = {2508.05961},
 primaryClass = {astro-ph.EP},
       adsurl = {https://ui.adsabs.harvard.edu/abs/2025AJ....170..257S}
}

@ARTICLE{Holland2002,
       author = {{Holland}, Heinrich D.},
        title = "{Volcanic gases, black smokers, and the great oxidation event}",
      journal = {\gca},
         year = 2002,
        month = nov,
       volume = {66},
       number = {21},
        pages = {3811-3826},
          doi = {10.1016/S0016-7037(02)00950-X},
       adsurl = {https://ui.adsabs.harvard.edu/abs/2002GeCoA..66.3811H}
}

@ARTICLE{Cooper1987,
       author = {{Cooper}, W.~J. and {Cooper}, D.~J. and {Saltzman}, E.~S. and {Mello}, W.~Z. de and {Savoie}, D.~L. and {Zika}, R.~G. and {Prospero}, J.~M.},
        title = "{Emissions of biogenic sulphur compounds from several wetland soils in Florida}",
      journal = {Atmospheric Environment},
         year = 1987,
        month = jan,
       volume = {21},
       number = {7},
        pages = {1491-1495},
          doi = {10.1016/0004-6981(87)90311-8},
       adsurl = {https://ui.adsabs.harvard.edu/abs/1987AtmEn..21.1491C}
}

@ARTICLE{AnejaCooper1989,
       author = {{Aneja}, Viney P. and {Cooper}, William J.},
        title = "{Biogenic Sulfur Emissions: A Review}",
      journal = {ACS Symposium Series},
         year = 1989,
        month = apr,
       volume = {393},
        pages = {2-13},
          doi = {10.1021/bk-1989-0393.ch001},
       adsurl = {https://ui.adsabs.harvard.edu/abs/1989ACSSS.393....2A}
}

@ARTICLE{Seager2013,
       author = {{Seager}, S. and {Bains}, W. and {Hu}, R.},
        title = "{Biosignature Gases in H$_{2}$-dominated Atmospheres on Rocky Exoplanets}",
      journal = {\apj},
         year = 2013,
        month = nov,
       volume = {777},
       number = {2},
          eid = {95},
        pages = {95},
          doi = {10.1088/0004-637X/777/2/95},
archivePrefix = {arXiv},
       eprint = {1309.6016},
 primaryClass = {astro-ph.EP},
       adsurl = {https://ui.adsabs.harvard.edu/abs/2013ApJ...777...95S}
}

@ARTICLE{DomagalGoldman2011,
       author = {{Domagal-Goldman}, Shawn D. and {Meadows}, Victoria S. and {Claire}, Mark W. and {Kasting}, James F.},
        title = "{Using Biogenic Sulfur Gases as Remotely Detectable Biosignatures on Anoxic Planets}",
      journal = {Astrobiology},
         year = 2011,
        month = jun,
       volume = {11},
       number = {5},
        pages = {419-441},
          doi = {10.1089/ast.2010.0509},
       adsurl = {https://ui.adsabs.harvard.edu/abs/2011AsBio..11..419D}
}

@ARTICLE{Pilcher2003,
       author = {{Pilcher}, Carl B.},
        title = "{Biosignatures of Early Earths}",
      journal = {Astrobiology},
         year = 2003,
        month = nov,
       volume = {3},
       number = {3},
        pages = {471-486},
          doi = {10.1089/153110703322610582},
       adsurl = {https://ui.adsabs.harvard.edu/abs/2003AsBio...3..471P}
}

@ARTICLE{Tsai2024,
       author = {{Tsai}, Shang-Min and {Innes}, Hamish and {Wogan}, Nicholas F. and {Schwieterman}, Edward W.},
        title = "{Biogenic Sulfur Gases as Biosignatures on Temperate Sub-Neptune Waterworlds}",
      journal = {\apjl},
         year = 2024,
        month = may,
       volume = {966},
       number = {2},
          eid = {L24},
        pages = {L24},
          doi = {10.3847/2041-8213/ad3801},
archivePrefix = {arXiv},
       eprint = {2403.14805},
 primaryClass = {astro-ph.EP},
       adsurl = {https://ui.adsabs.harvard.edu/abs/2024ApJ...966L..24T}
}

@ARTICLE{Lovelock1965,
       author = {{Lovelock}, J.~E.},
        title = "{A Physical Basis for Life Detection Experiments}",
      journal = {\nat},
         year = 1965,
        month = aug,
       volume = {207},
       number = {4997},
        pages = {568-570},
          doi = {10.1038/207568a0},
       adsurl = {https://ui.adsabs.harvard.edu/abs/1965Natur.207..568L}
}

@ARTICLE{KieneVisscher1987,
       author = {{Kiene}, Ronald P. and {Visscher}, Pieter T.},
        title = "{Production and Fate of Methylated Sulfur Compounds from Methionine and Dimethylsulfoniopropionate in Anoxic Salt Marsh Sediments}",
      journal = {Applied and Environmental Microbiology},
         year = 1987,
        month = oct,
       volume = {53},
       number = {10},
        pages = {2426-2434},
          doi = {10.1128/aem.53.10.2426-2434.1987},
       adsurl = {https://ui.adsabs.harvard.edu/abs/1987ApEnM..53.2426K}
}

@ARTICLE{HeinenLauwers1996,
       author = {{Heinen}, Wolfgang and {Lauwers}, Anne Marie},
        title = "{Organic sulfur compounds resulting from the interaction of iron sulfide, hydrogen sulfide and carbon dioxide in an anaerobic aqueous environment}",
      journal = {Origins of Life and Evolution of the Biosphere},
         year = 1996,
        month = apr,
       volume = {26},
       number = {2},
        pages = {131-150},
          doi = {10.1007/BF01809852},
       adsurl = {https://ui.adsabs.harvard.edu/abs/1996OLEB...26..131H}
}

@ARTICLE{Hanni2024,
       author = {{H{\"a}nni}, Nora and {Altwegg}, Kathrin and {Combi}, Michael and {Fuselier}, Stephen A. and {De Keyser}, Johan and {Ligterink}, Niels F.~W. and {Rubin}, Martin and {Wampfler}, Susanne F.},
        title = "{Evidence for Abiotic Dimethyl Sulfide in Cometary Matter}",
      journal = {\apj},
         year = 2024,
        month = nov,
       volume = {976},
       number = {1},
          eid = {74},
        pages = {74},
          doi = {10.3847/1538-4357/ad8565},
archivePrefix = {arXiv},
       eprint = {2410.08724},
 primaryClass = {astro-ph.EP},
       adsurl = {https://ui.adsabs.harvard.edu/abs/2024ApJ...976...74H}
}

@ARTICLE{Sans-Novo2025,
       author = {{Sanz-Novo}, Miguel and {Rivilla}, V{\'\i}ctor M. and {Endres}, Christian P. and {Lattanzi}, Valerio and {Jim{\'e}nez-Serra}, Izaskun and {Colzi}, Laura and {Zeng}, Shaoshan and {Meg{\'\i}as}, Andr{\'e}s and {L{\'o}pez-Gallifa}, {\'A}lvaro and {Mart{\'\i}nez-Henares}, Antonio and {San Andr{\'e}s}, David and {Tercero}, Bel{\'e}n and {de Vicente}, Pablo and {Mart{\'\i}n}, Sergio and {Requena-Torres}, Miguel A. and {Caselli}, Paola and {Mart{\'\i}n-Pintado}, Jes{\'u}s},
        title = "{On the Abiotic Origin of Dimethyl Sulfide: Discovery of Dimethyl Sulfide in the Interstellar Medium}",
      journal = {\apjl},
         year = 2025,
        month = feb,
       volume = {980},
       number = {2},
          eid = {L37},
        pages = {L37},
          doi = {10.3847/2041-8213/adafa7},
archivePrefix = {arXiv},
       eprint = {2501.08892},
 primaryClass = {astro-ph.GA},
       adsurl = {https://ui.adsabs.harvard.edu/abs/2025ApJ...980L..37S}
}

@ARTICLE{Reed2024,
       author = {{Reed}, Nathan W. and {Shearer}, Randall L. and {McGlynn}, Shawn Erin and {Wing}, Boswell A. and {Tolbert}, Margaret A. and {Browne}, Eleanor C.},
        title = "{Abiotic Production of Dimethyl Sulfide, Carbonyl Sulfide, and Other Organosulfur Gases via Photochemistry: Implications for Biosignatures and Metabolic Potential}",
      journal = {\apjl},
         year = 2024,
        month = oct,
       volume = {973},
       number = {2},
          eid = {L38},
        pages = {L38},
          doi = {10.3847/2041-8213/ad74da},
       adsurl = {https://ui.adsabs.harvard.edu/abs/2024ApJ...973L..38R}
}

@ARTICLE{RaulinToupance1975,
       author = {{Raulin}, F. and {Toupance}, G.},
        title = "{Effect of H$_{2}$S on the formation of organic compounds from C, N, H, S model atmospheres submitted to a glow discharge}",
      journal = {Origins of Life},
         year = 1975,
        month = oct,
       volume = {6},
       number = {4},
        pages = {507-512},
          doi = {10.1007/BF00928898},
       adsurl = {https://ui.adsabs.harvard.edu/abs/1975OrLi....6..507R}
}

@ARTICLE{Gao2016,
       author = {{Gao}, Connie W. and {Allen}, Joshua W. and {Green}, William H. and {West}, Richard H.},
        title = "{Reaction Mechanism Generator: Automatic construction of chemical kinetic mechanisms}",
      journal = {Computer Physics Communications},
         year = 2016,
        month = jun,
       volume = {203},
        pages = {212-225},
          doi = {10.1016/j.cpc.2016.02.013},
       adsurl = {https://ui.adsabs.harvard.edu/abs/2016CoPhC.203..212G}
}

@ARTICLE{Niraula2025,
       author = {{Niraula}, Prajwal and {de Wit}, Julien and {Hargreaves}, Robert and {Gordon}, Iouli E. and {Sousa-Silva}, Clara},
        title = "{The Detection-vs-Retrieval Challenge: Titan as an Exoplanet}",
      journal = {arXiv e-prints},
         year = 2025,
        month = jun,
          eid = {arXiv:2506.12144},
        pages = {arXiv:2506.12144},
          doi = {10.48550/arXiv.2506.12144},
archivePrefix = {arXiv},
       eprint = {2506.12144},
 primaryClass = {astro-ph.EP},
       adsurl = {https://ui.adsabs.harvard.edu/abs/2025arXiv250612144N}
}

@ARTICLE{MPI-Mainz2013,
       author = {{Keller-Rudek}, H. and {Moortgat}, G.~K. and {Sander}, R. and {S{\"o}rensen}, R.},
        title = "{The MPI-Mainz UV/VIS Spectral Atlas of Gaseous Molecules of Atmospheric Interest}",
      journal = {Earth System Science Data},
         year = 2013,
        month = dec,
       volume = {5},
       number = {2},
        pages = {365-373},
          doi = {10.5194/essd-5-365-201310.5194/essdd-6-411-2013},
       adsurl = {https://ui.adsabs.harvard.edu/abs/2013ESSD....5..365K}
}

@ARTICLE{Aaron2025,
       author = {{Werlen}, Aaron and {Dorn}, Caroline and {Burn}, Remo and {Schlichting}, Hilke E. and {Grimm}, Simon L. and {Young}, Edward D.},
        title = "{Sub-Neptunes Are Drier than They Seem: Rethinking the Origins of Water-rich Worlds}",
      journal = {\apjl},
         year = 2025,
        month = sep,
       volume = {991},
       number = {1},
          eid = {L16},
        pages = {L16},
          doi = {10.3847/2041-8213/adff73},
archivePrefix = {arXiv},
       eprint = {2507.00765},
 primaryClass = {astro-ph.EP},
       adsurl = {https://ui.adsabs.harvard.edu/abs/2025ApJ...991L..16W}
}

@ARTICLE{Innes2023,
       author = {{Innes}, Hamish and {Tsai}, Shang-Min and {Pierrehumbert}, Raymond T.},
        title = "{The Runaway Greenhouse Effect on Hycean Worlds}",
      journal = {\apj},
         year = 2023,
        month = aug,
       volume = {953},
       number = {2},
          eid = {168},
        pages = {168},
          doi = {10.3847/1538-4357/ace346},
archivePrefix = {arXiv},
       eprint = {2304.02698},
 primaryClass = {astro-ph.EP},
       adsurl = {https://ui.adsabs.harvard.edu/abs/2023ApJ...953..168I}
}

@ARTICLE{Leconte2024,
       author = {{Leconte}, J{\'e}r{\'e}my and {Spiga}, Aymeric and {Cl{\'e}ment}, No{\'e} and {Guerlet}, Sandrine and {Selsis}, Franck and {Milcareck}, Gwena{\"e}l and {Cavali{\'e}}, Thibault and {Moreno}, Rapha{\"e}l and {Lellouch}, Emmanuel and {Carri{\'o}n-Gonz{\'a}lez}, {\'O}scar and {Charnay}, Benjamin and {Lef{\`e}vre}, Maxence},
        title = "{A 3D picture of moist-convection inhibition in hydrogen-rich atmospheres: Implications for K2-18 b}",
      journal = {\aap},
         year = 2024,
        month = jun,
       volume = {686},
          eid = {A131},
        pages = {A131},
          doi = {10.1051/0004-6361/202348928},
archivePrefix = {arXiv},
       eprint = {2401.06608},
 primaryClass = {astro-ph.EP},
       adsurl = {https://ui.adsabs.harvard.edu/abs/2024A&A...686A.131L}
}

@ARTICLE{Jordan2025a,
       author = {{Jordan}, Sean and {Shorttle}, Oliver and {Rimmer}, Paul B.},
        title = "{Tracing the inner edge of the habitable zone with sulfur chemistry}",
      journal = {Science Advances},
         year = 2025,
        month = jan,
       volume = {11},
       number = {5},
          eid = {eadp8105},
        pages = {eadp8105},
          doi = {10.1126/sciadv.adp8105},
archivePrefix = {arXiv},
       eprint = {2501.17948},
 primaryClass = {astro-ph.EP},
       adsurl = {https://ui.adsabs.harvard.edu/abs/2025SciA...11P8105J}
}

@ARTICLE{Jordan2025b,
       author = {{Jordan}, Sean and {Shorttle}, Oliver and {Quanz}, Sascha P.},
        title = "{Planetary Albedo Is Limited by the Above-cloud Atmosphere: Implications for Sub-Neptune Climates}",
      journal = {\apj},
         year = 2025,
        month = nov,
       volume = {993},
       number = {1},
          eid = {86},
        pages = {86},
          doi = {10.3847/1538-4357/ae0192},
archivePrefix = {arXiv},
       eprint = {2504.12030},
 primaryClass = {astro-ph.EP},
       adsurl = {https://ui.adsabs.harvard.edu/abs/2025ApJ...993...86J}
}

@ARTICLE{Tsai2023,
       author = {{Tsai}, Shang-Min and {Lee}, Elspeth K.~H. and {Powell}, Diana and {Gao}, Peter and {Zhang}, Xi and {Moses}, Julianne and {H{\'e}brard}, Eric and {Venot}, Olivia and {Parmentier}, Vivien and {Jordan}, Sean and {Hu}, Renyu and {Alam}, Munazza K. and {Alderson}, Lili and {Batalha}, Natalie M. and {Bean}, Jacob L. and {Benneke}, Bj{\"o}rn and {Bierson}, Carver J. and {Brady}, Ryan P. and {Carone}, Ludmila and {Carter}, Aarynn L. and {Chubb}, Katy L. and {Inglis}, Julie and {Leconte}, J{\'e}r{\'e}my and {Line}, Michael and {L{\'o}pez-Morales}, Mercedes and {Miguel}, Yamila and {Molaverdikhani}, Karan and {Rustamkulov}, Zafar and {Sing}, David K. and {Stevenson}, Kevin B. and {Wakeford}, Hannah R. and {Yang}, Jeehyun and {Aggarwal}, Keshav and {Baeyens}, Robin and {Barat}, Saugata and {de Val-Borro}, Miguel and {Daylan}, Tansu and {Fortney}, Jonathan J. and {France}, Kevin and {Goyal}, Jayesh M. and {Grant}, David and {Kirk}, James and {Kreidberg}, Laura and {Louca}, Amy and {Moran}, Sarah E. and {Mukherjee}, Sagnick and {Nasedkin}, Evert and {Ohno}, Kazumasa and {Rackham}, Benjamin V. and {Redfield}, Seth and {Taylor}, Jake and {Tremblin}, Pascal and {Visscher}, Channon and {Wallack}, Nicole L. and {Welbanks}, Luis and {Youngblood}, Allison and {Ahrer}, Eva-Maria and {Batalha}, Natasha E. and {Behr}, Patrick and {Berta-Thompson}, Zachory K. and {Blecic}, Jasmina and {Casewell}, S.~L. and {Crossfield}, Ian J.~M. and {Crouzet}, Nicolas and {Cubillos}, Patricio E. and {Decin}, Leen and {D{\'e}sert}, Jean-Michel and {Feinstein}, Adina D. and {Gibson}, Neale P. and {Harrington}, Joseph and {Heng}, Kevin and {Henning}, Thomas and {Kempton}, Eliza M.-R. and {Krick}, Jessica and {Lagage}, Pierre-Olivier and {Lendl}, Monika and {Lothringer}, Joshua D. and {Mansfield}, Megan and {Mayne}, N.~J. and {Mikal-Evans}, Thomas and {Palle}, Enric and {Schlawin}, Everett and {Shorttle}, Oliver and {Wheatley}, Peter J. and {Yurchenko}, Sergei N.},
        title = "{Photochemically produced SO$_{2}$ in the atmosphere of WASP-39b}",
      journal = {\nat},
         year = 2023,
        month = may,
       volume = {617},
       number = {7961},
        pages = {483-487},
          doi = {10.1038/s41586-023-05902-2},
archivePrefix = {arXiv},
       eprint = {2211.10490},
 primaryClass = {astro-ph.EP},
       adsurl = {https://ui.adsabs.harvard.edu/abs/2023Natur.617..483T}
}

@ARTICLE{Rimmer2021,
       author = {{Rimmer}, Paul B. and {Jordan}, Sean and {Constantinou}, Tereza and {Woitke}, Peter and {Shorttle}, Oliver and {Hobbs}, Richard and {Paschodimas}, Alessia},
        title = "{Hydroxide Salts in the Clouds of Venus: Their Effect on the Sulfur Cycle and Cloud Droplet pH}",
      journal = {\psj},
         year = 2021,
        month = aug,
       volume = {2},
       number = {4},
          eid = {133},
        pages = {133},
          doi = {10.3847/PSJ/ac0156},
archivePrefix = {arXiv},
       eprint = {2101.08582},
 primaryClass = {astro-ph.EP},
       adsurl = {https://ui.adsabs.harvard.edu/abs/2021PSJ.....2..133R}
}

@ARTICLE{Rimmer2016,
       author = {{Rimmer}, P.~B. and {Helling}, Ch},
        title = "{A Chemical Kinetics Network for Lightning and Life in Planetary Atmospheres}",
      journal = {\apjs},
         year = 2016,
        month = may,
       volume = {224},
       number = {1},
          eid = {9},
        pages = {9},
          doi = {10.3847/0067-0049/224/1/9},
archivePrefix = {arXiv},
       eprint = {1510.07052},
 primaryClass = {astro-ph.EP},
       adsurl = {https://ui.adsabs.harvard.edu/abs/2016ApJS..224....9R}
}

@ARTICLE{Rimmer2019,
       author = {{Rimmer}, P.~B. and {Rugheimer}, S.},
        title = "{Hydrogen cyanide in nitrogen-rich atmospheres of rocky exoplanets}",
      journal = {\icarus},
         year = 2019,
        month = sep,
       volume = {329},
        pages = {124-131},
          doi = {10.1016/j.icarus.2019.02.020},
archivePrefix = {arXiv},
       eprint = {1902.08022},
 primaryClass = {astro-ph.EP},
       adsurl = {https://ui.adsabs.harvard.edu/abs/2019Icar..329..124R}
}

@ARTICLE{Harrison2025,
       author = {{Nicholls}, Harrison and {Pierrehumbert}, Raymond T. and {Lichtenberg}, Tim and {Soucasse}, Laurent and {Smeets}, Stef},
        title = "{Convective shutdown in the atmospheres of lava worlds}",
      journal = {\mnras},
         year = 2025,
        month = jan,
       volume = {536},
       number = {3},
        pages = {2957-2971},
          doi = {10.1093/mnras/stae2772},
archivePrefix = {arXiv},
       eprint = {2412.11987},
 primaryClass = {astro-ph.EP},
       adsurl = {https://ui.adsabs.harvard.edu/abs/2025MNRAS.536.2957N}
}

@ARTICLE{Dewitt2009,
       author = {{DeWitt}, H. Langley and {Trainer}, Melissa G. and {Pavlov}, Alex A. and {Hasenkopf}, Christa A. and {Aiken}, Allison C. and {Jimenez}, Jose L. and {McKay}, Christopher P. and {Toon}, Owen B. and {Tolbert}, Margaret A.},
        title = "{Reduction in Haze Formation Rate on Prebiotic Earth in the Presence of Hydrogen}",
      journal = {Astrobiology},
         year = 2009,
        month = jun,
       volume = {9},
       number = {5},
        pages = {447-453},
          doi = {10.1089/ast.2008.0289},
       adsurl = {https://ui.adsabs.harvard.edu/abs/2009AsBio...9..447D}
}

@ARTICLE{Park2017,
       author = {{Park}, Ki-Tae and {Jang}, Sehyun and {Lee}, Kitack and {Yoon}, Young Jun and {Kim}, Min-Seob and {Park}, Kihong and {Cho}, Hee-Joo and {Kang}, Jung-Ho and {Udisti}, Roberto and {Lee}, Bang-Yong and {Shin}, Kyung-Hoon},
        title = "{Observational evidence for the formation of DMS-derived aerosols during Arctic phytoplankton blooms}",
      journal = {Atmospheric Chemistry \& Physics},
         year = 2017,
        month = aug,
       volume = {17},
       number = {15},
        pages = {9665-9675},
          doi = {10.5194/acp-17-9665-2017},
       adsurl = {https://ui.adsabs.harvard.edu/abs/2017ACP....17.9665P}
}

@ARTICLE{Kevin2009,
       author = {{Zahnle}, K. and {Marley}, M.~S. and {Fortney}, J.~J.},
        title = "{Thermometric Soots on Warm Jupiters?}",
      journal = {arXiv e-prints},
         year = 2009,
        month = nov,
          eid = {arXiv:0911.0728},
        pages = {arXiv:0911.0728},
          doi = {10.48550/arXiv.0911.0728},
archivePrefix = {arXiv},
       eprint = {0911.0728},
 primaryClass = {astro-ph.EP},
       adsurl = {https://ui.adsabs.harvard.edu/abs/2009arXiv0911.0728Z}
}

@ARTICLE{Gao2017,
       author = {{Gao}, Peter and {Marley}, Mark S. and {Zahnle}, Kevin and {Robinson}, Tyler D. and {Lewis}, Nikole K.},
        title = "{Sulfur Hazes in Giant Exoplanet Atmospheres: Impacts on Reflected Light Spectra}",
      journal = {\aj},
         year = 2017,
        month = mar,
       volume = {153},
       number = {3},
          eid = {139},
        pages = {139},
          doi = {10.3847/1538-3881/aa5fab},
archivePrefix = {arXiv},
       eprint = {1701.00318},
 primaryClass = {astro-ph.EP},
       adsurl = {https://ui.adsabs.harvard.edu/abs/2017AJ....153..139G}
}

@ARTICLE{HarrisonAGNI,
       author = {{Nicholls}, Harrison and {Pierrehumbert}, Raymond and {Lichtenberg}, Tim},
        title = "{AGNI: A radiative-convective model for lava planet atmospheres}",
      journal = {The Journal of Open Source Software},
         year = 2025,
        month = may,
       volume = {10},
       number = {109},
          eid = {7726},
        pages = {7726},
          doi = {10.21105/joss.07726},
archivePrefix = {arXiv},
       eprint = {2506.00091},
 primaryClass = {astro-ph.IM},
       adsurl = {https://ui.adsabs.harvard.edu/abs/2025JOSS...10.7726N}
}

@article{blanskyBONDS,
author = {Blanksby, Stephen J. and Ellison, G. Barney},
title = {Bond Dissociation Energies of Organic Molecules},
journal = {Accounts of Chemical Research},
volume = {36},
number = {4},
pages = {255-263},
year = {2003},
doi = {10.1021/ar020230d},
    note ={PMID: 12693923},
URL = {https://doi.org/10.1021/ar020230d},
eprint = {https://doi.org/10.1021/ar020230d}

}

@ARTICLE{HuSeagerBains2013,
       author = {{Hu}, Renyu and {Seager}, Sara and {Bains}, William},
        title = "{Photochemistry in Terrestrial Exoplanet Atmospheres. II. H$_{2}$S and SO$_{2}$ Photochemistry in Anoxic Atmospheres}",
      journal = {\apj},
         year = 2013,
        month = may,
       volume = {769},
       number = {1},
          eid = {6},
        pages = {6},
          doi = {10.1088/0004-637X/769/1/6},
archivePrefix = {arXiv},
       eprint = {1302.6603},
 primaryClass = {astro-ph.EP},
       adsurl = {https://ui.adsabs.harvard.edu/abs/2013ApJ...769....6H}
}

@ARTICLE{Roy2025,
       author = {{Roy}, Pierre-Alexis and {Benneke}, Bj{\"o}rn and {Fournier-Tondreau}, Marylou and {Coulombe}, Louis-Philippe and {Piaulet-Ghorayeb}, Caroline and {Lafreni{\`e}re}, David and {Allart}, Romain and {Cowan}, Nicolas B. and {Dang}, Lisa and {Johnstone}, Doug and {Langeveld}, Adam B. and {Pelletier}, Stefan and {Radica}, Michael and {Taylor}, Jake and {Albert}, Lo{\"\i}c and {Doyon}, Ren{\'e} and {Flagg}, Laura and {Jayawardhana}, Ray and {MacDonald}, Ryan J. and {Turner}, Jake D.},
        title = "{Diversity in the haziness and chemistry of temperate sub-Neptunes}",
      journal = {Nature Astronomy},
         year = 2025,
        month = dec,
          doi = {10.1038/s41550-025-02723-3},
archivePrefix = {arXiv},
       eprint = {2512.10876},
 primaryClass = {astro-ph.EP},
       adsurl = {https://ui.adsabs.harvard.edu/abs/2025NatAs.tmp..256R}
}

@ARTICLE{Eddie2024,
       author = {{Schwieterman}, Edward W. and {Leung}, Michaela},
        title = "{An Overview of Exoplanet Biosignatures}",
      journal = {Reviews in Mineralogy and Geochemistry},
         year = 2024,
        month = jul,
       volume = {90},
       number = {1},
        pages = {465-514},
          doi = {10.2138/rmg.2024.90.13},
archivePrefix = {arXiv},
       eprint = {2404.15431},
 primaryClass = {astro-ph.EP},
       adsurl = {https://ui.adsabs.harvard.edu/abs/2024RvMG...90..465S}
}

@ARTICLE{Davenport2025,
       author = {{Davenport}, Brian and {Kempton}, Eliza M.-R. and {Nixon}, Matthew C. and {Ih}, Jegug and {Deming}, Drake and {Fu}, Guangwei and {May}, E.~M. and {Bean}, Jacob L. and {Gao}, Peter and {Rogers}, Leslie and {Malik}, Matej},
        title = "{TOI-421 b: A Hot Sub-Neptune with a Haze-free, Low Mean Molecular Weight Atmosphere}",
      journal = {\apjl},
         year = 2025,
        month = may,
       volume = {984},
       number = {2},
          eid = {L44},
        pages = {L44},
          doi = {10.3847/2041-8213/adcd76},
archivePrefix = {arXiv},
       eprint = {2501.01498},
 primaryClass = {astro-ph.EP},
       adsurl = {https://ui.adsabs.harvard.edu/abs/2025ApJ...984L..44D}
}

@INPROCEEDINGS{Louis2024,
       author = {{Maratrat}, Louis and {Carrasco}, Nathalie and {Chatain}, Audrey and {Buch}, Arnaud and {Vettier}, Ludovic and {Drant}, Thomas and {Jaziri}, Yassin and {Sohier}, Orianne and {Millan}, Maeva},
        title = "{Sulphur organic aerosols and their contribution to prebiotic chemistry}",
    booktitle = {European Planetary Science Congress},
         year = 2024,
        month = sep,
          eid = {EPSC2024-945},
        pages = {EPSC2024-945},
          doi = {10.5194/epsc2024-945},
       adsurl = {https://ui.adsabs.harvard.edu/abs/2024EPSC...17..945M}
}

@INPROCEEDINGS{Louis2025,
       author = {{Maratrat}, Louis and {Carrasco}, Nathalie and {Jaziri}, Yassin Adam and {Vettier}, Ludovic and {Millan}, Maeva},
        title = "{Deciphering the origin of organo-sulphur hazes: an exploration of the actual limits of sulphur chemistry with potential implications for prebiotic chemistry, the Early-Earth, and habitability of sulphur exoplanets}",
    booktitle = {EPSC-DPS Joint Meeting 2025},
         year = 2025,
       volume = {2025},
        month = sep,
          eid = {EPSC-DPS2025-326},
        pages = {EPSC-DPS2025-326},
          doi = {10.5194/epsc-dps2025-326},
       adsurl = {https://ui.adsabs.harvard.edu/abs/2025epsc.conf..326M}
}

@ARTICLE{France2016,
       author = {{France}, Kevin and {Loyd}, R.~O. Parke and {Youngblood}, Allison and {Brown}, Alexander and {Schneider}, P. Christian and {Hawley}, Suzanne L. and {Froning}, Cynthia S. and {Linsky}, Jeffrey L. and {Roberge}, Aki and {Buccino}, Andrea P. and {Davenport}, James R.~A. and {Fontenla}, Juan M. and {Kaltenegger}, Lisa and {Kowalski}, Adam F. and {Mauas}, Pablo J.~D. and {Miguel}, Yamila and {Redfield}, Seth and {Rugheimer}, Sarah and {Tian}, Feng and {Vieytes}, Mariela C. and {Walkowicz}, Lucianne M. and {Weisenburger}, Kolby L.},
        title = "{The MUSCLES Treasury Survey. I. Motivation and Overview}",
      journal = {\apj},
         year = 2016,
        month = apr,
       volume = {820},
       number = {2},
          eid = {89},
        pages = {89},
          doi = {10.3847/0004-637X/820/2/89},
archivePrefix = {arXiv},
       eprint = {1602.09142},
 primaryClass = {astro-ph.SR},
       adsurl = {https://ui.adsabs.harvard.edu/abs/2016ApJ...820...89F}
}

@ARTICLE{Youngblood2016,
       author = {{Youngblood}, Allison and {France}, Kevin and {Loyd}, R.~O. Parke and {Linsky}, Jeffrey L. and {Redfield}, Seth and {Schneider}, P. Christian and {Wood}, Brian E. and {Brown}, Alexander and {Froning}, Cynthia and {Miguel}, Yamila and {Rugheimer}, Sarah and {Walkowicz}, Lucianne},
        title = "{The MUSCLES Treasury Survey. II. Intrinsic LY{\ensuremath{\alpha}} and Extreme Ultraviolet Spectra of K and M Dwarfs with Exoplanets*}",
      journal = {\apj},
         year = 2016,
        month = jun,
       volume = {824},
       number = {2},
          eid = {101},
        pages = {101},
          doi = {10.3847/0004-637X/824/2/101},
archivePrefix = {arXiv},
       eprint = {1604.01032},
 primaryClass = {astro-ph.SR},
       adsurl = {https://ui.adsabs.harvard.edu/abs/2016ApJ...824..101Y}
}

@ARTICLE{Loyd2016,
       author = {{Loyd}, R.~O.~P. and {France}, Kevin and {Youngblood}, Allison and {Schneider}, Christian and {Brown}, Alexander and {Hu}, Renyu and {Linsky}, Jeffrey and {Froning}, Cynthia S. and {Redfield}, Seth and {Rugheimer}, Sarah and {Tian}, Feng},
        title = "{The MUSCLES Treasury Survey. III. X-Ray to Infrared Spectra of 11 M and K Stars Hosting Planets}",
      journal = {\apj},
         year = 2016,
        month = jun,
       volume = {824},
       number = {2},
          eid = {102},
        pages = {102},
          doi = {10.3847/0004-637X/824/2/102},
archivePrefix = {arXiv},
       eprint = {1604.04776},
 primaryClass = {astro-ph.SR},
       adsurl = {https://ui.adsabs.harvard.edu/abs/2016ApJ...824..102L}
}

@ARTICLE{Tang2008,
       author = {{Tang}, Yi-Zhen and {Pan}, Ya-Ru and {Sun}, Jing-Yu and {Sun}, Hao and {Wang}, Rong-Shun},
        title = "{Ab initio/DFT theory and multichannel RRKM study on the mechanisms and kinetics for the CH $_{3}$S + CO reaction}",
      journal = {Chemical Physics},
         year = 2008,
        month = mar,
       volume = {344},
       number = {3},
        pages = {221-226},
          doi = {10.1016/j.chemphys.2008.01.017},
       adsurl = {https://ui.adsabs.harvard.edu/abs/2008CP....344..221T}
}

@article{Yokota1979,
author = {Yokota, T. and Strausz, O. P.},
title = {Reaction of hydrogen atoms with dimethyl sulfide},
journal = {The Journal of Physical Chemistry},
volume = {83},
number = {25},
pages = {3196-3202},
year = {1979},
doi = {10.1021/j100488a003},

URL = { 
    
        https://doi.org/10.1021/j100488a003
    
    

},
eprint = { 
    
        https://doi.org/10.1021/j100488a003
    
    

}

}

@article{Ekwenchi1980,
author = {Ekwenchi, M. M. and Jodhan, A. and Strausz, O. P.},
title = {Reaction of hydrogen atoms with dimethyldisulfide},
journal = {International Journal of Chemical Kinetics},
volume = {12},
number = {6},
pages = {431-438},
doi = {https://doi.org/10.1002/kin.550120608},
url = {https://onlinelibrary.wiley.com/doi/abs/10.1002/kin.550120608},
eprint = {https://onlinelibrary.wiley.com/doi/pdf/10.1002/kin.550120608},
year = {1980}
}

@article{Zhu2006,
author = {Zhu, Li and Bozzelli, Joseph W.},
title = {Kinetics of the Multichannel Reaction of Methanethiyl Radical (CH3S•) with 3O2},
journal = {The Journal of Physical Chemistry A},
volume = {110},
number = {21},
pages = {6923-6937},
year = {2006},
doi = {10.1021/jp056209m},
    note ={PMID: 16722707},

URL = { 
    
        https://doi.org/10.1021/jp056209m
    
    

},
eprint = { 
    
        https://doi.org/10.1021/jp056209m
    
    

}

}

@article{Kerr2015,
author = {Kerr, Katherine E. and Alecu, Ionut M. and Thompson, Kristopher M. and Gao, Yide and Marshall, Paul},
title = {Experimental and Computational Studies of the Kinetics of the Reaction of Atomic Hydrogen with Methanethiol},
journal = {The Journal of Physical Chemistry A},
volume = {119},
number = {28},
pages = {7352-7360},
year = {2015},
doi = {10.1021/jp512966a},
    note ={PMID: 25872011},

URL = { 
    
        https://doi.org/10.1021/jp512966a
    
    

},
eprint = { 
    
        https://doi.org/10.1021/jp512966a
    
    

}

}

@article{Gersen2017,
author = {Gersen, Sander and van Essen, Martijn and Darmeveil, Harry and Hashemi, Hamid and Rasmussen, Christian Tihic and Christensen, Jakob Munkholdt and Glarborg, Peter and Levinsky, Howard},
title = {Experimental and Modeling Investigation of the Effect of H2S Addition to Methane on the Ignition and Oxidation at High Pressures},
journal = {Energy \& Fuels},
volume = {31},
number = {3},
pages = {2175-2182},
year = {2017},
doi = {10.1021/acs.energyfuels.6b02140},

URL = { 
    
        https://doi.org/10.1021/acs.energyfuels.6b02140
    
    

},
eprint = { 
    
        https://doi.org/10.1021/acs.energyfuels.6b02140
    
    

}

}

@article{Vijver2015,
author = {Van de Vijver, Ruben and Vandewiele, Nick M. and Bhoorasingh, Pierre L. and Slakman, Belinda L. and Seyedzadeh Khanshan, Fariba and Carstensen, Hans-Heinrich and Reyniers, Marie-Françoise and Marin, Guy B. and West, Richard H. and Van Geem, Kevin M.},
title = {Automatic Mechanism and Kinetic Model Generation for Gas- and Solution-Phase Processes: A Perspective on Best Practices, Recent Advances, and Future Challenges},
journal = {International Journal of Chemical Kinetics},
volume = {47},
number = {4},
pages = {199-231},
doi = {https://doi.org/10.1002/kin.20902},
url = {https://onlinelibrary.wiley.com/doi/abs/10.1002/kin.20902},
eprint = {https://onlinelibrary.wiley.com/doi/pdf/10.1002/kin.20902},
year = {2015}
}

@Article{Atkinson2004,
AUTHOR = {Atkinson, R. and Baulch, D. L. and Cox, R. A. and Crowley, J. N. and Hampson, R. F. and Hynes, R. G. and Jenkin, M. E. and Rossi, M. J. and Troe, J.},
TITLE = {Evaluated kinetic and photochemical data for atmospheric chemistry: Volume I - gas phase reactions of O$_{x}$, HO$_{x}$, NO$_{x}$ and SO$_{x}$ species},
JOURNAL = {Atmospheric Chemistry and Physics},
VOLUME = {4},
YEAR = {2004},
NUMBER = {6},
PAGES = {1461--1738},
URL = {https://acp.copernicus.org/articles/4/1461/2004/},
DOI = {10.5194/acp-4-1461-2004}
}

@article{Atkinson1997,
    author = {Atkinson, R. and Baulch, D. L. and Cox, R. A. and Hampson, R. F., Jr. and Kerr (Chairman), J. A. and Troe, J.},
    title = {Evaluated Kinetic and Photochemical Data for Atmospheric Chemistry: Supplement III. IUPAC Subcommittee on Gas Kinetic Data Evaluation for Atmospheric Chemistry},
    journal = {Journal of Physical and Chemical Reference Data},
    volume = {18},
    number = {2},
    pages = {881-1097},
    year = {1989},
    month = {04},
    issn = {0047-2689},
    doi = {10.1063/1.555832},
    url = {https://doi.org/10.1063/1.555832},
    eprint = {https://pubs.aip.org/aip/jpr/article-pdf/18/2/881/11744742/881_1_online.pdf},
}

@ARTICLE{Zhang2005,
       author = {{Zhang}, Qingzhu and {Sun}, Tingli and {Zhou}, Xuehua and {Wang}, Wenxing},
        title = "{Rate parameters and branching ratios for the multiple-channel reaction of dimethyl sulfide DMS with atomic H}",
      journal = {Chemical Physics Letters},
         year = 2005,
        month = oct,
       volume = {414},
       number = {4},
        pages = {316-321},
          doi = {10.1016/j.cplett.2005.08.084},
       adsurl = {https://ui.adsabs.harvard.edu/abs/2005CPL...414..316Z}
}

@article{Arthur1976,
  title={Reactions of methyl radicals. I. Hydrogen abstraction from dimethyl sulphide},
  author={Nl Arthur and M. Lee},
  journal={Australian Journal of Chemistry},
  year={1976},
  volume={29},
  pages={1483-1492},
  url={https://api.semanticscholar.org/CorpusID:102082476}
}

@article{Vidal2017,
author = { Vidal, T.~H.~G. and Loison, J.-C. and Jaziri, A.~Y. and Ruaud, M. and Gratier, P. and Wakelam, V. } ,
title = "{On the reservoir of sulphur in dark clouds: chemistry and elemental abundance reconciled}" ,
journal = {MNRAS} ,
year = 2017 ,
volume = 469 ,
pages = {435-447} ,
doi = {10.1093/mnras/stx828}
}

@ARTICLE{Pollack1993,
       author = {{Pollack}, Henry N. and {Hurter}, Suzanne J. and {Johnson}, Jeffrey R.},
        title = "{Heat flow from the earth's interior - Analysis of the global data set}",
      journal = {Reviews of Geophysics},
         year = 1993,
        month = aug,
       volume = {31},
       number = {3},
        pages = {267-280},
          doi = {10.1029/93RG01249},
       adsurl = {https://ui.adsabs.harvard.edu/abs/1993RvGeo..31..267P}
}

@ARTICLE{Ruiz2024,
       author = {{Ruiz}, Javier and {Jim{\'e}nez-D{\'\i}az}, Alberto and {Egea-Gonz{\'a}lez}, Isabel and {Romeo}, Ignacio and {Kirby}, Jon F. and {Audet}, Pascal},
        title = "{Heat loss and internal dynamics of Venus from lithosphere strength}",
      journal = {arXiv e-prints},
         year = 2024,
        month = jan,
          eid = {arXiv:2401.06558},
        pages = {arXiv:2401.06558},
          doi = {10.48550/arXiv.2401.06558},
archivePrefix = {arXiv},
       eprint = {2401.06558},
 primaryClass = {astro-ph.EP},
       adsurl = {https://ui.adsabs.harvard.edu/abs/2024arXiv240106558R}
}

@ARTICLE{Leung2025,
       author = {{Leung}, Michaela and {Tsai}, Shang-Min and {Schwieterman}, Edward W. and {Angerhausen}, Daniel and {Hansen}, Janina},
        title = "{Examining the Potential for Methyl Halide Accumulation and Detectability in Possible Hycean-type Atmospheres}",
      journal = {\apjl},
         year = 2025,
        month = mar,
       volume = {982},
       number = {1},
          eid = {L2},
        pages = {L2},
          doi = {10.3847/2041-8213/adb558},
archivePrefix = {arXiv},
       eprint = {2502.13856},
 primaryClass = {astro-ph.EP},
       adsurl = {https://ui.adsabs.harvard.edu/abs/2025ApJ...982L...2L}
}

@ARTICLE{Horst2018,
       author = {{H{\"o}rst}, Sarah M. and {He}, Chao and {Lewis}, Nikole K. and {Kempton}, Eliza M.-R. and {Marley}, Mark S. and {Morley}, Caroline V. and {Moses}, Julianne I. and {Valenti}, Jeff A. and {Vuitton}, V{\'e}ronique},
        title = "{Haze production rates in super-Earth and mini-Neptune atmosphere experiments}",
      journal = {Nature Astronomy},
         year = 2018,
        month = mar,
       volume = {2},
        pages = {303-306},
          doi = {10.1038/s41550-018-0397-0},
archivePrefix = {arXiv},
       eprint = {1801.06512},
 primaryClass = {astro-ph.EP},
       adsurl = {https://ui.adsabs.harvard.edu/abs/2018NatAs...2..303H}
}
\bibliographystyle{aasjournalv7}

\appendix

\section{Climate-chemistry calculation}

\textsc{Stand} provides the chemical reactions and reaction rate database for solving vertical atmospheric chemical profiles with the photochemical-kinetics code \textsc{Argo} \citep{Rimmer2016}. \textsc{Argo} is a 1D lagrangian code to calculate vertical atmospheric chemistry of planets by following a parcel of gas up and down through a column of the atmosphere iteratively, solving the time-dependent coupled continuity equation:
\begin{equation}
\dfrac{dn_{\rm X}}{dt} = P_{\rm X} - L_{\rm X} - \dfrac{\partial \Phi_{\rm X}}{\partial z},
\label{eq:argo_eqn}
\end{equation}
where, at a given altitude $z\,{\rm (cm)}$ (/pressure level) and time $t\,{\rm (s)}$, $n_{\ce{X}}$ (cm$^{-3}$) is the number density of species $\ce{X}$, $P_{\rm X}$ (cm$^{-3}$ s$^{-1}$) is the rate of production of species \ce{X}, $L_{\rm X}$ (s$^{-1}$) is the rate of loss of species \ce{X}, and $\partial \Phi_{\rm X}/\partial z$ (cm$^{-3}$ s$^{-1}$) represents the divergence of the vertical diffusion flux, encapsulating eddy-diffusion and molecular-diffusion. \textsc{Argo} is described in detail in \citet{Rimmer2016} and \citet{Rimmer2021}.

To solve the atmospheric chemistry, an initial compositional condition is provided at the base of the atmosphere, the stellar UV flux driving photochemistry is applied at the top of the atmosphere, and a temperature profile and eddy diffusion profile are provided as functions of altitude/pressure. The initial compositional conditions are applied as follows: in Section \ref{sec:compareDMS} the atmospheric mixing ratio of \ce{CH4} is 10\%, the mixing ratio of \ce{H2S} is varied between 1\,part per million (ppm) and 10\%, and the remainder is made up by \ce{H2}; in Sections \ref{sec:K2-18b} and \ref{sec:haze} the atmospheric mixing ratios are roughly those derived from the retrieval results of \citet{Hu2025} for K2-18b, with 10\% \ce{CH4}, 0.1\% \ce{CO2}, 10\% \ce{H2O}, and \ce{H2S} varied between 0.1\,ppm and 10\%, with the remainder made up by \ce{H2}. In all cases, the stellar UV flux used is that of GJ436 (M3.5, T$_{\rm eff}$\,=\,3416\,K) from the \textsc{Muscles} survey \citep{France2016,Youngblood2016,Loyd2016}.

In Section \ref{sec:compareDMS} the atmospheric chemistry is solved from 1\,bar to 10$^{-10}$\,bar for an isothermal atmosphere at 250\,K in order to test the relative efficiencies of pathways A and B for DMS and DMDS formation for reducing atmospheres generally. In Sections \ref{sec:K2-18b} and \ref{sec:haze} the atmospheric structure is solved more specifically for K2-18b, between 10$^{4}$\,bar and 10$^{-10}$ bar. To calculate the pressure-temperature profile of K2-18b we use the radiative-convective equilibrium code \textsc{Agni} \citep{HarrisonAGNI}. \textsc{Agni} calculates the 1D temperature structure and energy transport of an atmospheric column using the correlated-k method under the two-stream approximation. \textsc{Agni} employs the Newton-Raphson method to conserve energy fluxes through each level of the column while accounting for convection, condensation, and sensible heat transport. We use \textsc{Agni} here to calculate temperature profiles with the corresponding atmospheric mixing ratios from the retrievals of K2-18b's atmosphere from \citet{Hu2025} and the inferred estimate of 10\% \ce{H2O} mixing ratio, while allowing \ce{H2O} to condense out from the observable upper atmosphere during the calculation of the temperature structure. In Section \ref{sec:haze} we explore a range of interior heat flux boundary conditions from 0\,Wm$^{-2}$ up to 1\,Wm$^{-2}$, encompassing the heat fluxes of modern Earth (87\,mWm$^{-2}$, \citet{Pollack1993}) and Venus (31\,mWm$^{-2}$, \citet{Ruiz2024}), and approaching the heat flux of Io ($\sim$1.5\,--\,4\,Wm$^{-2}$). For an interior heat flux of 0\,Wm$^{-2}$, the temperature approaches a 1050\,K isotherm in the deep atmosphere which would be too low to melt a silicate mantle. For an interior heat flux of 0.01\,Wm$^{-2}$ a temperature of 1600\,K is achieved at 10$^{5}$\,bar pressure and a molten silicate mantle would be possible. The temperature at an atmosphere-magma ocean interface then increases with increasing interior heat flux. The pressure level where water condenses out and the atmospheric temperature structure above are each independent of our choice of interior heat flux.



\end{document}